\documentclass[10pt,letterpaper]{article}
\usepackage[top=0.85in,left=2.75in,footskip=0.75in]{geometry}

\usepackage{amsmath,amssymb}
\newcommand{\xnew}[1]{\textcolor{blue}{#1}}

\usepackage{changepage}

\usepackage[utf8x]{inputenc}

\usepackage{textcomp,marvosym}

\usepackage{cite}

\usepackage{nameref,hyperref}

\usepackage[right]{lineno}

\usepackage{microtype}
\DisableLigatures[f]{encoding = *, family = * }

\usepackage[table]{xcolor}

\usepackage{array}

\newcolumntype{+}{!{\vrule width 2pt}}

\newlength\savedwidth

\raggedright
\usepackage[aboveskip=1pt,labelfont=bf,labelsep=period,justification=raggedright,singlelinecheck=off]{caption}

\makeatletter
\renewcommand{\@biblabel}[1]{\quad#1.}
\makeatother

\date{}

\usepackage{lastpage,fancyhdr,graphicx}
\usepackage{epstopdf}
\fancyheadoffset[L]{2.25in}
\fancyfootoffset[L]{2.25in}
\begin{document}
\vspace*{0.2in}

\begin{flushleft}
{\Large
\textbf\newline{\textbf{Enhanced storage capacity with errors in scale-free Hopfield neural networks: an analytical study}} 
}
\newline
\\
Do-Hyun Kim\textsuperscript{1*},
Jinha Park\textsuperscript{2},
Byungnam Kahng\textsuperscript{2*}
\\
\bigskip
\textsuperscript{1} Department of Physics, Sogang University, Seoul 04107, Korea
\\
\textsuperscript{2} CCSS,  CTP and  Department of Physics and Astronomy, Seoul National University, Seoul 08826, Korea
\\
\bigskip

%
%





* Corresponding authors \\
E-mail: dohyunkim@sogang.ac.kr(D-HK) or bkahng@snu.ac.kr(BK)

\end{flushleft}
\section*{Abstract}
The Hopfield model is a pioneering neural network model with associative memory retrieval. The analytical solution of the model in mean field limit revealed that memories can be retrieved without any error up to a finite storage capacity of $O(N)$, where $N$ is the system size. Beyond the threshold, they are completely lost. Since the introduction of the Hopfield model, the theory of neural networks has been further developed toward realistic neural networks using analog neurons, spiking neurons, etc. Nevertheless, those advances are based on fully connected networks, which are inconsistent with recent experimental discovery that the number of connections of each neuron seems to be heterogeneous, following a heavy-tailed distribution. Motivated by this observation, we consider the Hopfield model on scale-free networks and obtain a different pattern of associative memory retrieval from that obtained on the fully connected network: the storage capacity becomes tremendously enhanced but with some error in the memory retrieval, which appears as the heterogeneity of the connections is increased. Moreover, the error rates are also obtained on several real neural networks and are indeed similar to that on scale-free model networks.



\section*{Introduction}

Human neuroscience has attracted increasing attention through various studies. Among such activities, the retrieval or recall of associative memory in neural networks is \xnew{a} historically noticeable issue~\cite{Josselyn15, Amit89}.
Associative memory means the ability to link, learn and remember the relationship between independent and unrelated items such as one man's name (e.g., Albert Einstein) and his previous achievement (e.g., $E=mc^{2}$).
Neural network models of associative memory have been used to explain how the brain stores and recalls  long-term memories.
These models incorporate the so-called Hebbian rule for a cell assembly, a group of excitatory neurons mutually coupled by strong synapses~\cite{Hebb}:
Memory storage occurs when a cell assembly is created by Hebbian synaptic plasticity, and memory retrieval or recall occurs when the neurons in the cell assembly are activated by a stimulus.
Neural network models of associative memory assume that information exists alternatively as neural activity or as synaptic connections.
When novel information first enters into the brain, it is encoded in a pattern of neural activity. If this information is stored as memory, the neural activity leaves synaptic connections modified.
The stored information can be retrieved when the modified connections again become active~\cite{Kandel13}.

A Hopfield model was introduced in the year 1982~\cite{Hopfield}, following which there have been enormous researches based on the premise that retrieval of associative memory occurs by way of pattern recognition in collective excitations of associated neurons. The Hopfield model has been thus accepted as a paradigmatic neural network model for associative memory retrieval. A Hopfield network is composed of Ising-type neurons with two discrete states, that is, an excitation pattern of each neuron is in a state either $+1$ or $-1$, representing excited and rest states for transmitting or not transmitting a signal, respectively. Each neuron is supposed to be connected to all other neurons and then the synaptic weight is updated by the Hebbian rule. An associative memory was introduced as an activity pattern by collective excitations of the associated neurons. A simple example of excitation updated by the Hebbian rule is presented in Section 1 of the supporting information (S1 File).

In this model study, a mathematical quantity called {\sl energy} was introduced to each memory pattern. This quantity decreases as the retrieval activity proceeeds until the system reaches a stable state, at which the retrieved pattern is consistent to a stored pattern. We will show why such behavior occurs in the Hopfield model in Section 2 of the S1 File.
This behavior is similar to the dynamic process of a thermodynamic system toward an equilibrium state, at which a free energy becomes minimum.

In associative neural networks, the number of patterns that can be stored in synaptic connections is a central quantity to calibrate their performance. The maximal number of patterns that can be stored before having total confusion divided by the total number of neurons in the network is called the \textit{storage capacity} of the network. If the cell assemblies share no neurons in common, the number of patterns that can be stored is as many as the total number of neurons in the network. If the cell assemblies share some neurons, however, interference may occur among those cell assemblies. If too many patterns are stored in the network, the stored patterns can be interfered and the quality of memory recall can be lowered. Therefore, interference effect induces errors in memory retrieval and reduces the retrievability~\cite{Kandel13}.
Furthermore, stochastic processes in any real network system can occur. We thus consider a temperature $T$, which is not to be understood as a physical temperature, but as a noise strength for the stochastic process.  In the limit of zero temperature ($T=0$), the deterministic model is recovered.
The Hopfield model treated such noise effect in terms of temperature, and invoke the formalism of Boltzmann statistical mechanics to obtain various properties of the retrievability as a function of both the number of stored patterns and  temperature, i.e., noise strength. Amit \textit{et al.} calculated the storage capacity of the Hopfield model on a fully connected network at finite temperature using a method in statistical physics~\cite{Amit85,Amit87}.

Recent experimental results using the functional magnetic resonance imaging (fMRI) technique have revealed that functional neural networks in resting state are not fully connected networks but the number of connections of each coarse-grained neuron are heterogeneous following a heavy-tailed distribution~\cite{Bullmore09,Heuvel13}. For further discussion, the number of connections of a neuron is referred to as degree using a term in graph theory. Even though it is not clear yet how functional connections of an individual neuron is related to its anatomical connections, it becomes more acceptable that neurons with similar patterns of connection tend to exhibit similar functions~\cite{Bullmore09}. This suggests that neural networks need not necessarily be a fully connected network. Indeed, there exists supporting evidences: a structural neural network of the worm {\it Caenorhabditis elegans} has a power-law tail in the degree distribution~\cite{Barabasi99}. Mathematically, this is expressed as $P_{d}(k)\sim k^{-\gamma}$, where $P_d(k)$ is the degree distribution, $k$ denotes degree and $\gamma$ is the degree exponent. Such networks are referred to as scale-free (SF) networks. Recent electrophysiological data~\cite{Song05} also revealed that the distribution of synaptic connections strength follows a log-normal distribution, which has a heavy tail similar to that in the power-law distribution. Moreover, it is known that the brain damages of schizophrenia~\cite{Rubinov13,Steullet14} or comatose patients~\cite{Achard12} are caused by the malfunctioning of hub neurons. Thus, hub neurons that have many connections is likely to exist in a brain network.

This paper aims to investigate the properties of the retrieval patterns created by the Hopfield model on SF networks at finite temperature analytically. We also compare obtained results with previous numerical results obtained from fully connected networks and diverse SF networks such as the Barab\'{a}si-Albert model with $\gamma = 3$~\cite{Stauffer03} and the Molloy-Reed model with several values of $\gamma$~\cite{Torres04}. We also use the Chung-Lu  model~\cite{ChungLu,Cho09} to construct uncorrelated SF networks over the entire range of $\gamma$. The details of the Chung-Lu model are presented in the Method section. Particularly we consider the limit $\gamma\to 2$, which is the case observed in the fMRI data~\cite{Eguiluz05,Heuvel08,Gallos12}.

\begin{figure*}
\centering
\includegraphics[width=0.88\textwidth]{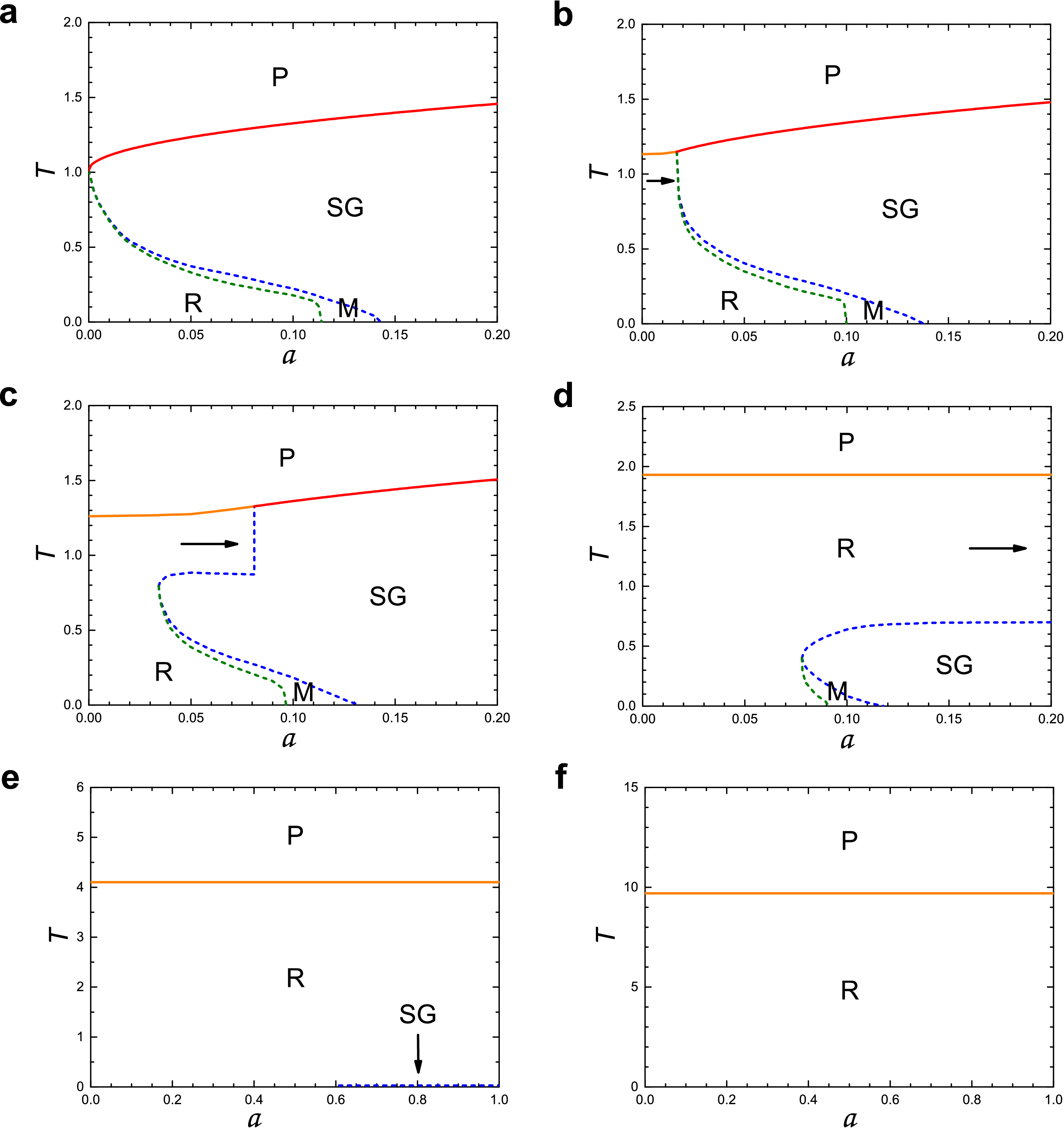}
\caption{{\bf Phase diagram of the Hopfield model in the plane of $(T, a)$.} Here $T$ and $a$ denote temperature and storage rate, respectively. Degree exponent $\gamma$ is infinity in {\bf a}, 5.0 in {\bf b,} 4.0 in {\bf c,} 3.0 in {\bf d}, 2.04 in {\bf e,} and 2.01 in {\bf f}.
{\bf P} represents the paramagnetic phase, in which $m = 0$, $q = 0$, and $r=0$ because of thermal fluctuations. Here, $m$, $q$, and $r$ are given by Eqs.~(29-31) of the S1 File, respectively. {\bf SG} does the spin-glass phase, in which $m = 0, q > 0$, and $r > 0$. In the P and SG phases, the retrieval of stored patterns is impossible. Thus, they are often referred to as the confusion phase. The retrieval phase is denoted as {\bf R}, in which $m > 0, q > 0$, and $r > 0$. The retrieval of stored memory is possible. Finally, {\bf M} does the mixed phase, in which the features of both the retrieval and the spin-glass phases coexist.
As the degree exponent $\gamma$ is decreased from infinity in {\bf a} through $\gamma=2.01$ in {\bf f}, the retrieval phase not only intrudes into the region of the SG phase, but also raises the boundary of the phase {\bf P} to a higher temperature region. Eventually the SG phase remains on the $T=0$ axis when $\gamma = \gamma_{c} \simeq 2.04$, in which the phase {\bf R} spans most of the low-temperature region. Thus, memory retrieval is enhanced. The phase boundary was obtained by performing numerical calculations for the Chung-Lu SF networks with the system size $N = 1000$ and mean degree $K = 5.0$. Solid and dotted lines or curves indicate the second-order and first-order transitions, respectively. We note that the case {\bf a} on ER network is nearly the same as that in mean field limit obtained in the original Hopfield model. }
\label{fig:fig1}
\end{figure*}

The main results of our studies are presented in Figs.~\ref{fig:fig1} and \ref{fig:fig2}. In Fig~\ref{fig:fig1}, we present properties of the retrieval pattern for various degree exponents $\gamma$ as a fuction of temperature $T$ and storage rate $a$. When $\gamma <  2.04$ (Fig~\ref{fig:fig1}(e) and (f)), the retrieval phase spans most of the low-temperature (noise) region; thus memory retrieval in the system is appreciably enhanced compared with the one of the original Hopfield model~\cite{Amit85,Amit87}. In Fig~\ref{fig:fig2}, the error rate, a fraction of neurons which fails memory retrieval, is obtained as a function of storage rate $a\equiv p/N$, the number of stored patterns per neuron, at zero temperature $T = 0$. Remarkably when $\gamma = 2.01$, the error rate is almost zero when the number of stored patterns $p$ is small, and  gradually increases but is less than 0.3 even when $p$ is increased to the total number of neurons $N$. This implies that the storage capacity becomes tremendously enhanced as the SF network becomes extremely heterogeneous in structure, but there occur some errors. We remark that the previous solution of the Hopfield model on fully connected networks~\cite{Amit85,Amit87} revealed that the error rate is zero up to a certain threshold, but beyond which it becomes one-half and the system falls into a total confusion state. Based on these results, we think that the solution of the Hopfield model on SF networks reflects more a normal brain in the point that it provides the case of imperfect memory retrieval with some error even when its storage capacity is small. Moreover, the  result is timely in accord with a recent experimental discovery that storage capacity of brain is in the petabyte range, as much as entire web, ten times more than previously thought~\cite{Bartol15}. We hope that our result will provide some theoretical development for modeling associative memory networks in neuroscience.

\section*{Results}
\subsection*{Model Development}
\subsubsection*{Hopfield model on a given network}
In the Hopfield model, each neuron at node $i$ of a given neural network (denoted as $G$) has an Ising spin $S_i$ with two states, $S_i=+1$ and $S_i=-1$ representing excited and rest states for transmitting or not transmitting a signal, respectively~\cite{Hopfield}. The Hamiltonian (corresponding to energy of the system) is introduced as
\begin{equation}
\mathcal{H} = -\sum_{(i,j) \in G} J_{ij} S_{i} S_{j},
\label{eq:HopfieldHamiltonian}
\end{equation}
where the coupling strength $J_{ij}$ between two connected nodes $(i,j)\in G$, known as the synapse efficacy, takes the Hebbian form,
\begin{equation}
J_{ij} = \frac{1}{K} \sum_{\mu=1}^{p} \xi_{i}^{\mu} \xi_{j}^{\mu}.
\label{eq:HopfieldSynapse}
\end{equation}
$K$ is the mean degree, i.e., the average number of edges of the network $G$ of size $N$. $\xi_{i}^{\mu}$ is another quantity assigned to node $i$, which also has either $+1$ or $-1$.
A collective quantity $\{\xi_{i}^{\mu}\}$ represents a memory pattern denoted by $\mu$ that is stored in the system. The index $\mu$ runs $\mu= 1, \dots, p$, which means that the number of memory patterns stored is $p$. Whereas $\xi_{i}^{\mu}$ is fixed throughout the dynamics. Starting from some initial values of $\{S_i(t=0)\}$, the state of each spin is updated asynchronously as
\begin{equation}
S_{i} (t+1) = {\rm sgn} \bigg(\sum_{j} J_{ij} S_{j}(t)\bigg).
\label{eq:HopfieldUpdate}
\end{equation}
When $\sum_j J_{ij} S_j(t)$ becomes zero, $S_i(t+1)=+1$ is assigned definitely.

As the updating is repeated, the energy given by  (\ref{eq:HopfieldHamiltonian}) reduces and the system reaches a local minumum state. In this state, each spin $S_i$ becomes equivalent to $\xi_i^{\mu}$ for a given pattern $\mu$, and the stored memory is retrieved. The underlying mechanism for such converging behavior is explained in Section 2 of the S1 File. In real-world systems, however, the repeated dynamics may not be deterministic as Eq.~\ref{eq:HopfieldUpdate}, but it may include some noise. To take into account of noise effect, the original study\cite{Amit85} of Hopfield model invoked the formalism developed in equilibrium statistical mechanics at finite temperature, in which temperature represents noise strength.

\subsubsection*{Ensemble average of the Hopfield model over different network configurations and stored patterns}

Thus far, we consider the Hopfield model on a given network. However, connection profiles of SF networks can be different from sample to sample even though they follow the same degree distribution. Thus, we need to take average of physical quantities over the ensemble of different network configurations. To proceed, we consider the probability that a given SF network $G$ exists in the ensemble. That is given as
\begin{align}
P_{K}(G) &= \prod_{(i,j) \in G} f_{ij} \prod_{(i,j) \notin G} (1 - f_{ij}),
\end{align}
where $f_{ij}$ is the probability to connect a link between two nodes $i$ and $j$. It was derived that $f_{ij} = 1 - \exp (-N K w_{i} w_{j})$ \cite{ChungLu,Cho09}. The factor $w_{i}$ is a weight of node $i$ reflecting heterogeneous degrees of a SF network. The explicit form of the weight factor is presented in the Method section. Then the ensemble average over different networks for any given physical quantity $A$ is taken as
\begin{equation}
\langle A \rangle_{K} = \sum_{G} P_{K}(G) A(G),
\label{eq:A_K}
\end{equation}
where $\langle \cdots \rangle_{K}$ denotes the average over different graph configurations. $A(G)$ represents any physical quantity obtained in a graph $G$.

Next, we consider a situation in which stored patterns are not deterministic, but stochastically generated. This case is considered for the purpose of testing the efficiency of the algoritumc. Specifically, each pattern $\mu$ is created with the population of $\xi_i^{\mu}=1$ with probability $1/2$ and that of $\xi_i^{\mu}=-1$ with probability $1/2$ for each node $i$. Then, the probability that a pattern $\mu$ is created is given as
\begin{equation}
P(\{\xi_{i}^\mu\}) = \prod_{i=1}^{N} \Big[ \frac{1}{2} \delta (\xi_{i}^{\mu} - 1) +  \frac{1}{2} \delta (\xi_{i}^{\mu} + 1) \Big].
\end{equation}
We take the average of any given physical quantity $A$ over the ensemble of different stored patterns as
\begin{equation}
\langle A \rangle_{\xi} = \int d\xi_{i}^{\mu} P(\{\xi_{i}^{\mu}\}) A(\{\xi_{i}^{\mu}\}) ,
\label{eq:A_xi}
\end{equation}
where $\langle \cdots \rangle_{\xi}$ is an average over the quenched disorder of $\xi_{i}^{\mu}$.

\begin{figure*}
\centering
\includegraphics[width=0.88\textwidth]{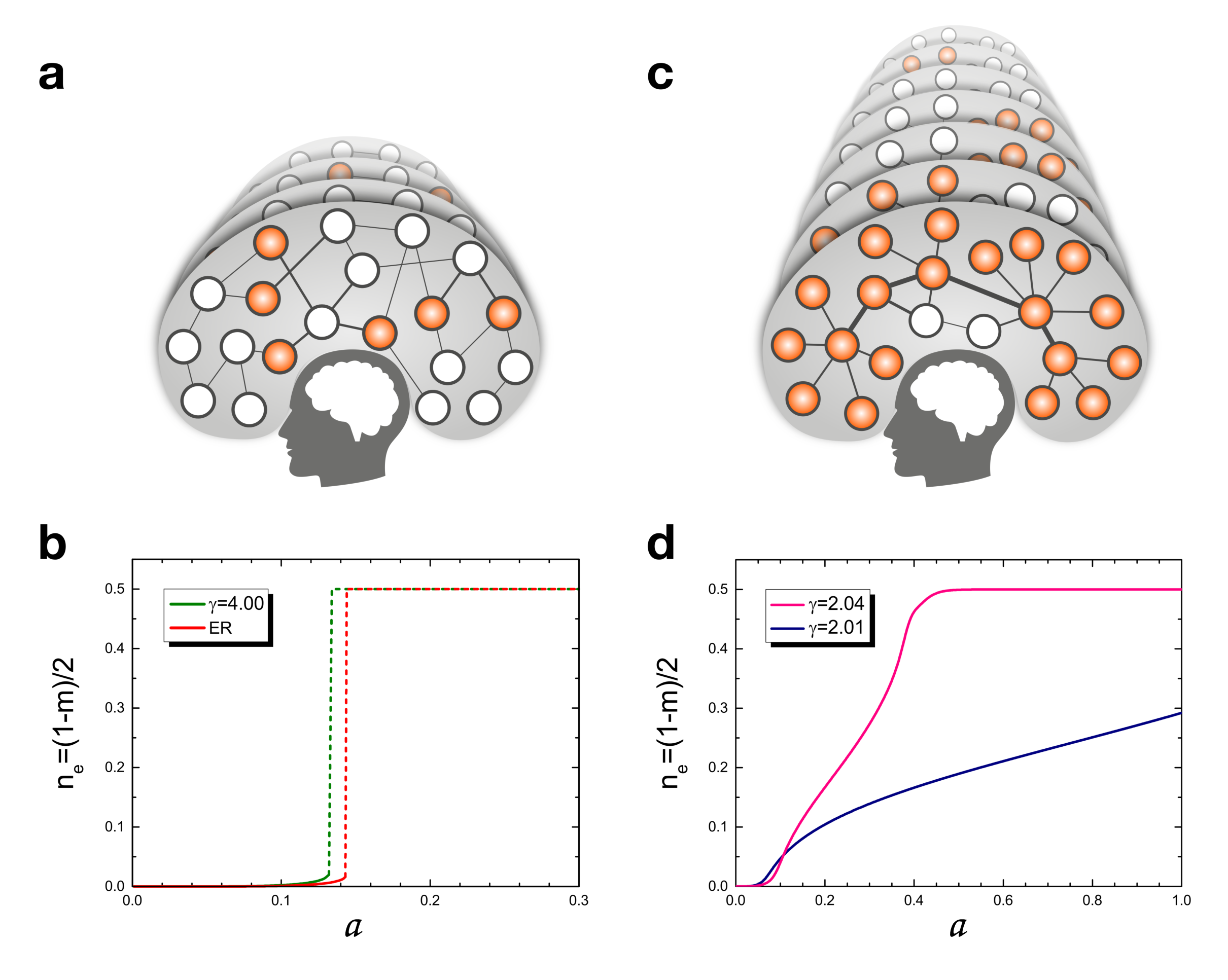}
\caption{{\bf Conceptual figures of the storage capacities and the error rates.} {\bf a} for an ER random network and {\bf c} for a SF network. {\bf b} and {\bf d} Plot of the error rate $n_e\equiv (1-m)/2$ vs storage rate $a$ for several $\gamma$ values of the Chung-Lu model at $T=0$. Here, numerical values are obtained using $N = 1000$ and $K = 5.0$. The dotted lines for $\gamma \gg 2.0$ indicate the sudden jumps from small error rates to the state of $n_{e}=0.5$. ({\bf a} and {\bf c}, Figure courtesy of Joonwon Lee)}
\label{fig:fig2}
\end{figure*}

\subsection*{Analytic solutions}

\subsubsection*{The order parameters}
We characterize the phases of the Hopfield model by three quantities, which are often called the order parameters in statistical physics as follows:
i) The overlap parameter defined as $m_{\alpha}^{\mu} \equiv \sum_{i}^{N} w_{i} \langle \xi_{i}^{\mu} S_{i}^{\alpha} \rangle$, which represents the extent of which the $\mu$-th pattern of memory $\xi^{\mu}$ and the $\alpha$-th state of the system $S^{\alpha}$ overlap with each other. Thus, this quantity measures the retrieval success rate of the $\mu$-th memory. We remark that the factor $w_{i}$ is required to take into account heterogeneous degrees. $w_{i}$ is proportional to the degree $k_{i}$ of neuron $i$~\cite{Goh01,Lee06,Kim05,Kim14}. Actually this weight $w_i$ is the same as that used in the Chung-Lu model introduced in \nameref{methods:ChungLu}.
ii) The so-called spin glass order parameter $q_{\alpha \beta}  \equiv \sum_{i}^{N} w_{i} \langle S_{i}^{\alpha}  S_{i}^{\beta} \rangle$ representing the extent of which the two states $\alpha$ and $\beta$ of the replica overlap each other. The replica is a quantity introduced in the spin-glass theory to resolve the technical difficulties in taking the ensemble average $\langle \cdots \rangle_K$ and $\langle \cdots \rangle_{\xi}$ introduced in Eqs.~(\ref{eq:A_K}) and (\ref{eq:A_xi}).
iii) A new quantity is introduced as $r_{\alpha \beta}\equiv (N/p) \sum_{\mu= 2}^{p}  m_{\alpha}^{\mu} m_{\beta}^{\mu} $. This quantity is necessary to derive the first two order parameters, and is interpreted as the sum of the effects of each non-retrieval pattern.

Next, we take the replica-symmetric solution by setting $q_{\alpha \beta}=q$ and $r_{\alpha \beta} =r$ for all $\alpha \neq \beta$ and $m_{\alpha}^{1}=m$ for all $\alpha$.
Then the order parameters $m$, $q$ and $r$ at zero temperature are obtained as
\begin{eqnarray}
m &=& \sum_{i=1}^{N} w_{i} ~\mathrm{erf} \Big( \sqrt{\frac{N w_{i}}{2 a r}} m \Big) \label{eq8}\\
q &=& 1 - \frac{1}{N K} \sum_{i=1}^{N} \sqrt{\frac{2 N w_{i}}{\pi a r}} ~ \exp \Big(- \frac{N w_{i}}{2 a r} m^{2} \Big) \label{eq9}\\
r &=& \frac{q}{ (1- K + K q)^{2}}, \label{eq10}
\end{eqnarray}
which are used for obtaining the error rate.
Here, storage rate $a$ is defined as the number of existing memory patterns $p$ divided by the total number of neurons in a given network, i.e., the network size $N$. Therefore, the storage capacity $a_{c}$ is the maximum value of storage rate $a$. Detailed calculations for those parameters are presented in Section 4 of the S1 File.

\subsubsection*{Phase diagram and error rates}

The phase diagram we obtained are shown in Fig~\ref{fig:fig1} in the $T-a$ plane for different degree exponents $\gamma$ but with fixed $N = 1000$ and $K = 5.0$. First, in Fig~\ref{fig:fig1}(a), we consider the case of the Erd\H{o}s and R\'enyi (ER) network, equivalent to the limit $\gamma \to \infty$. This phase diagram is nearly the same as that obtained on fully connected network in Ref.~\cite{Amit87}. There exist four different phases: i) The paramagnetic phase denoted as P, in which $m=0$, $q=0$, and $r=0$ because of thermal fluctuations. ii) The spin-glass phase denoted as SG, in which $m=0$, $q > 0$, and $r > 0$. In those phases P and SG, the retrieval of stored patterns is impossible. Thus, they are referred to as the confusion phases. iii) The retrieval phase denoted as R, in which $m> 0$, $q > 0$, and $r > 0$, and in which the retrieval of stored memory is possible. Finally, the mixed phase denoted as M, in which the properties of both the retrieval and the spin-glass phases coexist.

The free energy of the system in each phase has also been investigated in Ref.~\cite{Amit87}. In phase R, a stored pattern, for instance, $\{\xi_i^{\mu}\}_i$ is matched to a state of the system $\{S_{i}^{\alpha}\}_i$ at the global minimum of free energy. However, in phase M, a stored pattern is matched to a state of the system in a metastable state, while the SG state lies at the global minimum.

For $a = 0$, which occurs when $p\sim \mathcal{O}(N)$ in the limit $N\to \infty$, there exist two phases, P and R.
As $a$ is increased slightly from $a=0$, the SG and M phases appear between the phases P and  R. The transition between the P and the SG phases is a second-order transition, whereas that between SG and M is a first-order transition. Likewise, the transition between M and R is also first order. At $T=0$, the transition between R and M occurs at $a_{m} \simeq 0.114$, and the transition between M and SG occurs at $a_{c} \simeq 0.143$. Therefore, as temperature is lowered from a sufficiently high value, successive transitions occur   following the steps P $\to$ SG $\to$ M $\to$ R for $a < a_{m}$. For $a_{m} < a < a_{c}$, successive  transitions occur following the steps P $\to$ SG $\to$ M, and for $a > a_{c}$, a transition from P $\to$ SG occurs.

Second, we obtain the phase diagram for finite $\gamma$ values in Fig~\ref{fig:fig1}(b)-(e). That undergoes drastic changes depending on $\gamma$. Remarkably, the R phase intrudes into the region of the SG phase, but it also raises the boundary of the P phase to a higher-temperature region. As $\gamma$ approaches 2.04 in Fig~\ref{fig:fig1}(e), we observe that the R phase prevails, whereas the SG phase shrinks until it only exists at $T=0$ in the region $a > 0.6$. Such changes in the $T-a$ diagrams can be understood analytically by examining the phase boundary between the P and SG phases. The details are presented in Section 5 of the S1 File.

Third, we consider a particular case, the noiseless case $T=0$ in Fig~\ref{fig:fig2}. The order parameters $m$, $q$ and $r$ are given in Eqs.~(\ref{eq8}-\ref{eq10}).
The figures show the behavior of the error rates $n_{e}$ as a function of $a$ with $N = 1000$ and $K = 5.0$.
The error rate means the relative error of the neural networks and is defined as $n_{e} \equiv (1-m)/2$.
This figure is obtained analytically for various degree exponent values including $\gamma \simeq 2.0$. We first consider the case in which $\gamma \to \infty$, i.e., the ER limit.
The result of this case is almost the same as that obtained in Ref.~\cite{Amit87}.  When $a$ is less than $a_{c} \simeq 0.143$, $n_{e}$ is very small, such that the error rate is negligible and the system is almost in the error-free state. The obtained value $a_{c}$ is close to  $a_{c} \simeq 0.138$, which was obtained on the fully connected network in Ref.~\cite{Amit87}.  As $a$ reaches $a_{c}$, $n_{e}$ suddenly jumps to 0.5 as shown in Fig~\ref{fig:fig2}(b). This means that for $a > a_{c}$, the system is in the error-full state (i.e., the state of complete confusion).
As $\gamma$ decreases, this behavior persists up to $\gamma_{c}\simeq 2.7$, and no longer holds for $\gamma \le \gamma_{c}$. Note that the value of $a_{c}$ slightly decreases with decreasing $\gamma$, but their dependences are almost negligible.

Next, when $\gamma$ approaches 2.0, $n_{e}$ noticeably changes from the step-function-like shape to a monotonously increasing one as shown in Fig~\ref{fig:fig2}(d). As $\gamma$ is lowered further and approaches $2.0$, the range of $a$ for the state of complete confusion, with $n_{e}=0.5$, disappears and $a_{c}$ cannot be defined anymore. For instance, at $\gamma = 2.01$, the error rate becomes less than $0.3$ for the entire range of $a$. Therefore, when $\gamma$ is lowered to as small as 2.0, while the range of $a$ for the state of $n_{e}=0$ (the state of perfect retrieval without error) is reduced, the range of $a$ for the state of $n_{e}=0.5$, the region of the state of complete confusion disappears. The details are presented in Section 6 of the S1 File.
These behaviors have never been observed yet in previous studies~\cite{Stauffer03,Torres04,Castillo04}. Fig~\ref{fig:fig2} thus implies that as the network changes from the ER network to the extremely heterogeneous SF network with degree exponent two, the storage capacity becomes tremendously enhanced, but some error occurs. This result suggests that hubs play central roles in memory retrieval.


\section*{Conclusion and Discussion}

We obtained the results that as the network changes from a hub-absent network to a SF network with degree exponent just above two, the storage capacity becomes tremendously enhanced, but some error occurs. These features seem to be in accordance with what we experience in everyday life and with a recent discovery of enormously high storage capability in the human brain. Thus hubs, i.e., neurons with a large number of synapses, and other nodes with heterogeneous degrees in neural networks play a central role in enhancing the storage capacity. It is interesting that a normal human brain has such a structure, even though detailed structural properties such as modularity and degree correlation are not yet known. We consider the effect of degree correlation near $\gamma=2.0$ by performing a similar analysis of the static model~\cite{Goh01,Lee06,Kim05,Kim14},  the degree correlation of which is disassortative between $2 < \gamma < 3$.  We find that correlated SF networks near $\gamma = 2.0^+$ have lower values of $n_{e}$ than the uncorrelated ones. The details are presented in Section 8 of the S1 File.
We also checked the error rate $n_e$ on several real neural networks, having heavy-tailed degree distributions but with undetermined degree exponents due to the small network size. The obtained patterns of the error rates are indeed similar to the theoretical prediction. The details are presented in Section 9 of the S1 File.

Our results might provide some guidelines for constructing an artificial neural network, provided that it is constructed on the basis of the Hopfield model: If we want to construct an artificial neural network that is capable of perfect memory recall with a large value of $a_{c}$, as a basic topology of the artificial neural network, it may be more appropriate to choose an ER-type random network or a fully connected one. However, if we prefer to construct an artificial neural network that supports an extended range of storage rate and tolerate a small range of errors, it may be more appropriate to choose an SF-type network with the degree exponent $\gamma \simeq 2.0$.
In this case, we can construct an artificial neural network with relatively low cost compared with a fully-connected one which needs very many synapses of $O(N^{2})$.\\

\section*{Methods}
\label{methods:ChungLu}
\subsection*{Construction of scale-free networks}
To construct uncorrelated SF networks, we use the Chung-Lu (CL) model \cite{ChungLu, Cho09}:
We start with a fixed number of $N$ vertices. Each vertex $i$ $(i=1,2,\ldots, N)$ is assigned a weight
\begin{eqnarray}
w_{i}=\frac{(i + i_{0} - 1)^{- \nu}}{\sum_{j=1}^{N} (j + i_{0} - 1)^{- \nu}},
\end{eqnarray}
where $\nu$ is a control parameter in the range $[0,1)$, and $i_{0}$ is constant given by
\begin{eqnarray}
i_{0} = \left\{
\begin{array}{ll}
[10 \sqrt{2} (1 - \nu)]^{1/\nu} N^{1 - 1/2\nu} & (1/2 < \nu < 1), ~~~\\
1  & (0 \leq \nu < 1/2).
\end{array}\right.
\end{eqnarray}

A pair of vertices $(i, j)$ is chosen with the probabilities $w_{i}$ and $w_{j}$, respectively, and they are connected with an edge, unless the pair is already connected. This process is repeated $NK/2$ times. Then, the resulting network becomes an uncorrelated SF network following a power-law degree distribution, $P_{d} (k) \sim k^{-\gamma}$, where $k$ denotes the degree and $\gamma$ is the degree exponent with $\gamma = 1 + 1/\nu$.
In such random networks, the probability that a given pair of vertices $(i, j)$ ($i \neq j$) is not connected by an edge, as denoted by $1-f_{ij}$, is given by $(1-2w_{i}w_{j})^{NK/2} \simeq \exp (-NK w_{i}w_{j})$, while the connection probability $f_{ij}=1-\exp (-N K w_{i}w_{j})$.

As a particular case, when we choose $i_{0}$ as 1 for all the values of $\nu$, the CL model reduces to the static model \cite{Goh01, Lee06, Kim05, Kim14}, which has correlations for the range of $1/2 < \nu < 1$ \cite{Goh01, Lee06}. Therefore, the weights for the static model are given by
\begin{eqnarray}
w_{i}=\frac{i^{-\nu}}{\zeta_{N}(\nu)}
\end{eqnarray}
where $\nu$ is a control parameter in the range $[0,1)$, and $\zeta_{N}(\nu) \equiv \sum_{j=1}^{N} j^{-\nu} \simeq {N^{1-\nu}}/(1-\nu)$.
Note that $f_{ij} \simeq N K w_{i} w_{j}$ for finite $K$, however, $f_{ij} \simeq 1$ for $2 < \gamma < 3$ and $ij \ll N^{3-\gamma}$.

For the Erd\H{o}s-R\'enyi (ER) graph \cite{Erdos59,Erdos60,Bollobas01}, $\nu$ becomes 0 and the weights of both models become $w_{i}=1/N$,
independent of the index $i$. Since $w_{i} w_{j}=1/N^{2}$, the fraction of bonds present becomes
$f_{ij} \simeq K/N$ and the total number of the connected edges $L$ is $NK/2$. So $K$ becomes the mean degree in the ER graph.

When $K$ approaches $N$, this network becomes a fully-connected one or a regular lattice with infinite-range interaction. Note that previous studies on the Hopfield model focused mainly on such extreme cases of $K \to N$ \cite{Amit85, Amit87, Amit89, Mezard, Nishimori}.




\section*{Acknowledgments}
This work was supported by the NRF of Korea (Grant Nos. 2014R1A3A2069005 and 2015R1A5A7037676) and Sogang University (Grant Nos. 201610033.01 and 201710066.01). We would like to express our appreciation toward S.-H. Lee and Joonwon Lee.

\newpage

{\Large
\textbf\newline{\textbf{Supporting information for ``Enhanced storage capacity with errors in scale-free Hopfield neural networks: an analytical study"}} 
}


\section{Hopfield model simulation on a random network}
\label{appen:simulation}

In order to give an insight for the Hopfield model simulation, we performed a simulation following the random updating method of the Hopfield model on a random network:
An excitation pattern of a random network is denoted by $\{ \xi_{i}^{\mu} \}$, where $i$  is the index of the neuron, $\mu (= 1, \dots, p)$  is the index of the excitation pattern, and $p$  is the number of existing patterns, equivalent to the number of stored memories. The value of $\xi_{i}^{\mu}$  is either $+1$ or $-1$.
An initial spin configuration of each neural network is set to be the first pattern, i.e., $S_i(t=1)=\xi_i^{\mu=1}$ for each node $i$. Once this spin configuration is set up, at the next step $t=2$, a neuron $i$ is randomly chosen, its Ising spin state is updated asynchronously as
\begin{eqnarray}
S_{i} (t+1) = {\rm sgn} \Big( \sum_{j\in {\rm n.n. of } i} J_{ij} S_{j}(t) \Big),
\label{eq:HopfieldUpdate2}
\end{eqnarray}
where the synapse efficacy $J_{ij}$ is taken as $J_{ij} = (1/K) \sum_{\mu=1}^{p} \xi_{i}^{\mu} \xi_{j}^{\mu}$ following the Hebb's rule, and $K(\equiv 2L/N)$ is the mean degree of a given random network. If $\sum_jJ_{ij}S_j(t)$ becomes zero, then $S_i(t+1)$ takes +1 definitely. This updating is repeated until the system reaches a stable fixed point, which is supposed to be the energy minimum state of the system. Thus the retrieved  pattern is the spin configuration in the energy minimum state. This whole procedure completes one run.

An example of the Hopfield dynamics with a random network of size $N=9$ and the number of edges $L=15$ is shown in Fig. 3. Its adjacency matrix $\hat A$ is given as
\begin{eqnarray}
{\hat A} = \left( \begin{array}{ccccccccc}
0 & 1 & 1 & 0 & 1 & 1 & 0 & 0 & 0 \\
1 & 0 & 1 & 1 & 1 & 0 & 1 & 0 & 1 \\
1 & 1 & 0 & 1 & 0 & 0 & 1 & 1 & 0 \\
0 & 1 & 1 & 0 & 0 & 0 & 0 & 0 & 0 \\
1 & 1 & 0 & 0 & 0 & 1 & 0 & 1 & 0 \\
1 & 0 & 0 & 0 & 1 & 0 & 0 & 0 & 0 \\
0 & 1 & 1 & 0 & 0 & 0 & 0 & 0 & 1 \\
0 & 0 & 1 & 0 & 1 & 0 & 0 & 0 & 0 \\
0 & 1 & 0 & 0 & 0 & 0 & 1 & 0 & 0 \\
\end{array} \right).
\end{eqnarray}
The two patterns $\xi^1=\{1, -1, 1, 1, 1, 1, 1, -1, 1\}=\textrm{``H''}$ and $\xi^2=\{-1, 1, -1, -1, 1, -1, -1, 1, -1\}=\textrm{``I''}$ are encoded to the intercoupling strengths $J$, following the Hebbian rule, Eq.~(2) of the main paper.
\begin{eqnarray}
J &=& \frac{1}{K}{\hat A}\otimes\big((\xi^1)^T\xi^1 + (\xi^2)^T \xi^2\big) \\
&=&\frac{3}{5}{\hat A}\otimes
\left( \begin{array}{ccccccccc}
1 & -1 & 1 & 1 & 1 & 1 & 1 & -1 & 1 \\
-1 & 1 & -1 & -1 & -1 & -1 & -1 & 1 & -1 \\
1 & -1 & 1 & 1 & 1 & 1 & 1 & -1 & 1 \\
1 & -1 & 1 & 1 & 1 & 1 & 1 & -1 & 1 \\
1 & -1 & 1 & 1 & 1 & 1 & 1 & -1 & 1 \\
1 & -1 & 1 & 1 & 1 & 1 & 1 & -1 & 1 \\
1 & -1 & 1 & 1 & 1 & 1 & 1 & -1 & 1 \\
-1 & 1 & -1 & -1 & -1 & -1 & -1 & 1 & -1 \\
1 & -1 & 1 & 1 & 1 & 1 & 1 & -1 & 1 \\
\end{array} \right) \nonumber \\
&+&\frac{3}{5}{\hat A}\otimes
\left( \begin{array}{ccccccccc}
1 & -1 & 1 & 1 & -1 & 1 & 1 & -1 & 1 \\
-1 & 1 & -1 & -1 & 1 & -1 & -1 & 1 & -1 \\
1 & -1 & 1 & 1 & -1 & 1 & 1 & -1 & 1 \\
1 & -1 & 1 & 1 & -1 & 1 & 1 & -1 & 1 \\
-1 & 1 & -1 & -1 & 1 & -1 & -1 & 1 & -1 \\
1 & -1 & 1 & 1 & -1 & 1 & 1 & -1 & 1 \\
1 & -1 & 1 & 1 & -1 & 1 & 1 & -1 & 1 \\
-1 & 1 & -1 & -1 & 1 & -1 & -1 & 1 & -1 \\
1 & -1 & 1 & 1 & -1 & 1 & 1 & -1 & 1 \\
\end{array}
\right)\\
&=& \left( \begin{array}{ccccccccc}
0 & -\frac{6}{5} & \frac{6}{5} & \frac{6}{5} & 0 & 0 & 0 & 0 & 0 \\
-\frac{6}{5} & 0 & -\frac{6}{5} & 0 & 0 & 0 & -\frac{6}{5} & \frac{6}{5} & 0 \\
\frac{6}{5} & -\frac{6}{5} & 0 & \frac{6}{5} & 0 & \frac{6}{5} & 0 & -\frac{6}{5} & 0 \\
\frac{6}{5} & 0 & \frac{6}{5} & 0 & 0 & \frac{6}{5} & \frac{6}{5} & 0 & \frac{6}{5} \\
0 & 0 & 0 & 0 & 0 & 0 & 0 & 0 & 0 \\
0 & 0 & \frac{6}{5} & \frac{6}{5} & 0 & 0 & 0 & 0 & \frac{6}{5} \\
0 & -\frac{6}{5} & 0 & \frac{6}{5} & 0 & 0 & 0 & 0 & 0 \\
0 & \frac{6}{5} & -\frac{6}{5} & 0 & 0 & 0 & 0 & 0 & 0 \\
0 & 0 & 0 & \frac{6}{5} & 0 & \frac{6}{5} & 0 & 0 & 0 \\
\end{array} \right),
\end{eqnarray}
where $\otimes$ denotes an elementwise multiplication between two matrices and the mean degree $K=\frac{5}{3}$.
Now starting from an initial condition $S(t=0)=\{1, -1, -1, 1, 1, -1, 1, -1, 1\}$, the system evolves under Eq.~\eqref{eq:HopfieldUpdate2} and reaches a stationary state $\xi^1=\textrm{``H''}$.
Also from an initial condition $S(t=0)=\{1, 1, -1, -1, 1, -1, -1, 1, 1\}$, the system evolves under Eq.~\eqref{eq:HopfieldUpdate2} and reaches a stationary state $\xi^2=\textrm{``I''}$. See Fig. 3(c) and (d).

\section{Hebbian rule}
\label{appen:energy}
\textbf{1) The system's energy decreases in time.}\newline
(i) Suppose $S_k$ was flipped at time $t$, $S_k(t+1) = -S_k(t)$ and $S_i(t+1) = S_i(t)$ for all other indices $i\neq k$.
This implies,
$$\textrm{sgn}\left(\sum_{i}J_{ik}S_i(t)\right)\neq\textrm{sgn} \left(S_k(t)\right).$$
Then,
\begin{align*}
H(t+1) - H(t) &= -\sum_{ij} J_{ij} S_i(t+1)S_j(t+1) + \sum_{ij} J_{ij} S_i(t)S_j(t)\\
&= -2 \sum_{i\neq k} J_{ik}S_i(t+1)S_k(t+1)\\
&=2\left(\sum_{i\neq k} J_{ik}S_i(t))\right)S_k(t) < 0
\end{align*}
(ii) Suppose if $S_k$ was not flipped, i.e. $\vec{S}(t+1)=\vec{S}(t)$. Thus $H(t+1)-H(t)=0$.

Therefore with the asynchronous sgn updating rule, the energy of the system decreases monotonically. $H(t+1)-H(t)\leq 0$.
\newline

\noindent \textbf{2) By the Hebbian encoding, each stored pattern corresponds to a local energy minimum.}\newline
Let $S=\xi^1+\delta S$.
\begin{align*}
H[S] &= -\sum_{ij} J_{ij}S_iS_j =-\sum_{ij}J_{ij}(\xi^1+\delta S)_i(\xi^1+\delta S)_j \\
\frac{\delta H}{\delta S}\bigg|_{S=\xi^1} & = -2 \sum_{ij} J_{ij} \xi_i^1 =-\frac{2}{N}\sum_{i\neq j}\sum_{\mu=1}^{p}\xi_i^\mu\xi_j^\mu\xi_i^1 \\
&= -\sum_{\mu}\Big(\sum_j\xi_j^\mu\Big)\bigg(\frac{1}{N}\sum_i\xi_i^\mu\xi_i^1\bigg) +\frac{2}{N}\sum_\mu\Big(\sum_i\xi_i^1\Big)=0.
\end{align*}
where $\sum_j\xi_j^\mu=0$, because the stored patterns are assumed unbiased, and $\frac{1}{N}\sum_i \xi_i^\mu\xi_i^1 = 0$ because the stored patterns $\xi^\mu,\xi^\nu,\cdots$ are assumed uncorrelated.
\begin{align*}
\frac{\delta^2 H}{\delta S^2}\bigg|_{S=\xi^1} &= - \sum_{ij} J_{ij} \\
&= -\frac{1}{N}\sum_{i\neq j}\sum_{\mu=1}^{p}\xi_i^\mu\xi_j^\mu \\
&= -\frac{1}{N}\sum_{\mu}\Big(\sum_i \xi_i^\mu\Big)^2 + \frac{1}{N}\sum_{\mu}\sum_i\big(\xi_i^\mu\big)^2 = p > 0.
\end{align*}
The argument also holds for $\xi^2, \xi^3, \cdots \xi^p$. Therefore each pattern $\xi^\mu$ is a local energy minimum. The Hebbian rule shapes the energy function to become quadratic near each pattern $\xi^\mu$. For random networks uncorrelated with patterns, $\hat{A}_{ij}$ randomly sets some of the terms to zero and argument 2) is effectively unchanged.

From the above arguments 1) and 2), we conclude that the stored patterns $\xi^\mu$ are stationary states of the Hopfield dynamics. However, these states are not the only stationary states. Some other local energy minima can exist, which are states usually given as superpositions of the existing patterns. In the SG regime, the number of local energy minima become much larger than the number of stored patterns, hindering the memory retrieval process.

\section{Solution of the Hopfield model on scale-free networks}
\label{appen:hopfield}

Now we consider the Hopfield model on SF networks. Since Amit \textit{et al.} successfully applied replica analysis of spin glass theory to the Hopfield model \cite{Amit85, Amit87, Amit89}, neural networks have been regarded as analogous systems of spin glasses (SGs) \cite{Mezard, Nishimori}. The SG transitions in Euclidean space have already been studied by means of various theoretical methods \cite{Mezard, Nishimori}. Most of such studies have been carried out on regular lattices or for the infinite-range interaction model on fully connected graphs. To study the SG transitions on SF networks, we followed the previous approach used for the dilute Ising SG model with infinite-range interactions, i.e., the Ising SG model on the ER graph, first performed by Viana and Bray
\cite{Viana85,Kanter87,Mezard87,Mottishaw87,Wong88,Monasson98}, and applied it successfully to the static model of SF networks \cite{Kim05}. The Ghatak-Sherrington SG model was also studied on the same networks \cite{Kim14}.

In the spin glass and its analogous systems, the replica method is conventionally used to evaluate the free energy~\cite{Mezard,Nishimori} to calculate the ensemble averages in $-\beta F = \langle \langle \ln Z \rangle_{\xi} \rangle_{K}$, where $Z$ is the partition function for a given distribution of $\{\xi_{i}\}$ and $\{K\}$ on a particular graph $G$ and $\beta=1/T$. Here, once $\xi$ and $K$ are given, they become fixed throughout all retrieval processes, thus they are regarded as quenched variables in statistical physics. For such systems with the quenched variables, we cannot calculate $\langle \langle \ln Z \rangle_{\xi} \rangle_{K}$ mathematically. Instead, we can carry out such calculations with the so-called replica method, i.e., the mathematical relation $X^{n} = e^{n
\ln X} \approx 1+n \ln X$ as $n \to 0$.
Therefore, using the replica method, we obtain the following relation: $-\beta F = \langle \langle \ln Z \rangle_{\xi} \rangle_{K}= \lim_{n \to 0} [\langle \langle Z^{n} \rangle_{\xi} \rangle_{K}-1]/n$.
By evaluating the $n$-th power of the partition function $\langle\langle Z^{n} \rangle_{\xi} \rangle_{K}$, we finally obtain the free energy. For simplicity, only the first pattern $(\mu=1)$ is retrieved. This simplification is based on the fact that the replica symmetry only appears to be broken in a particular case at $T=0$~\cite{Amit89}.

Using the replica method, we evaluate the $n$-th power of the partition function $Z^{n}$,
\begin{eqnarray}
\langle \langle Z^{n} \rangle_{\xi} \rangle_{K} &=&
\textrm{Tr} ~ \Big\langle \Big\langle \exp \Big( \frac{\beta}{K}
\sum_{ (ij) \in G} \sum_{\mu=1}^{p} \sum_{\alpha}^{n} \xi_{i}^{\mu} \xi_{j}^{\mu} S_{i}^{\alpha} S_{j}^{\alpha} \Big)
\Big\rangle_{\xi} \Big\rangle_{K} \nonumber \\
&=& \textrm{Tr} ~ \exp \Big[ \sum_{i < j} \ln \Big\{ 1 + f_{ij} \Big( \Big\langle \exp \Big( \frac{\beta}{K}
\sum_{\mu=1}^{p} \sum_{\alpha}^{n} \xi_{i}^{\mu} \xi_{j}^{\mu} S_{i}^{\alpha}S_{j}^{\alpha} \Big)
\Big\rangle_{\xi} -1 \Big) \Big\} \Big],
\end{eqnarray}
where the trace $\textrm{Tr}$ is taken over all replicated spins $S_{i}^{\alpha} (= \pm 1)$, and $\alpha=1,\ldots,n$ is the replica index.
Using the relation,
\begin{eqnarray}
\Big\langle \exp \Big( \frac{\beta}{K} \sum_{\mu=1}^{p} \sum_{\alpha}^{n} \xi_{i}^{\mu} \xi_{j}^{\mu} S_{i}^{\alpha} S_{j}^{\alpha} \Big) \Big\rangle_{\xi}
= \Big\langle \prod_{\mu=1}^{p} \prod_{\alpha}^{n} \Big[ \cosh (\beta/K) \Big(1 + \xi_{i}^{\mu} \xi_{j}^{\mu} S_{i}^{\alpha} S_{j}^{\alpha} \tanh (\beta/K) \Big) \Big] \Big\rangle_{\xi},~
\end{eqnarray}
$\langle \langle Z^{n} \rangle_{\xi} \rangle_{K}$ term becomes
\begin{eqnarray}
\langle \langle Z^{n} \rangle_{\xi} \rangle_{K} &\stackrel{n \to 0}{=}&
\textrm{Tr} ~ \prod_{i \neq j} \exp \Big[ \frac{1}{2} \Big\langle
\Big[ NK w_{i} w_{j} + NK \mathbf{T}_{1} w_{i} \xi_{i}^{1} S_{i}^{\alpha} w_{j} \xi_{j}^{1} S_{j}^{\alpha} + \cdots \Big] \times \nonumber \\
& &\prod_{\mu=2}^{p}\Big[ 1 + \mathbf{T}_{1} \xi_{i}^{\mu} S_{i}^{\alpha} \xi_{j}^{\mu} S_{j}^{\alpha} + \cdots \Big] \Big\rangle_{\xi} - NK w_{i} w_{j} \Big] \nonumber\\
&=& \textrm{Tr} ~ \prod_{i \neq j} \exp \frac{1}{2} \Big\langle
NK \mathbf{T}_{1} (w_{i} \xi_{i}^{1} S_{i}^{\alpha}) (w_{j} \xi_{j}^{1} S_{j}^{\alpha})
+ \sum_{\mu=2}^{p} NK \mathbf{T}_{1} (w_{i} \xi_{i}^{\mu} S_{i}^{\alpha}) (w_{j} \xi_{j}^{\mu} S_{j}^{\alpha}) \nonumber\\
&&
+ NK \mathbf{T}_{2} (w_{i} \xi_{i}^{1} S_{i}^{\alpha} \xi_{i}^{1} S_{i}^{\beta}) (w_{j} \xi_{j}^{1} S_{j}^{\alpha} \xi_{j}^{1} S_{j}^{\beta}) \nonumber \\
&&
+ \sum_{\mu=2}^{p}  NK \mathbf{T}_{2} (w_{i} \xi_{i}^{\mu} S_{i}^{\alpha} \xi_{i}^{\mu} S_{i}^{\beta}) (w_{j} \xi_{j}^{\mu} S_{j}^{\alpha} \xi_{j}^{\mu} S_{j}^{\beta}) + \cdots \Big\rangle_{\xi},~
\end{eqnarray}
where
\begin{eqnarray}
\mathbf{T}_{l}(T) &\equiv& \cosh^n (\beta/K) \tanh^{l} (\beta/K) \stackrel{n \to 0}{\longrightarrow} \tanh^{l} (\beta/K) \nonumber \\
&& (l=1,2,\ldots).
\end{eqnarray}
Because all $p$ stored patterns were generally randomly, we can examine the retrieval pattern of the first pattern ($\mu = 1$) only. Thus, we separate the contribution of the first pattern from the rest. The $\langle \langle Z^{n} \rangle_{\xi} \rangle_{K}$ term can be linearized using the Hubbard-Stratonovich transformation \cite{Stratonovich,Hubbard}
\begin{eqnarray}
\exp\{\frac{1}{2}\lambda a^{2}\}=\sqrt{\frac{\lambda}{2 \pi}}\int_{-\infty}^{\infty} dx ~ \exp \Big\{-\frac{1}{2}\lambda x^{2}+ \lambda a x \Big\}~.
\end{eqnarray}
Therefore,
\begin{eqnarray}
\langle \langle Z^{n} \rangle_{\xi} \rangle_{K} &=&
\textrm{Tr} ~ \int \prod_{\mu}^{p} \prod_{\alpha}^{n} dm_{\alpha}^{\mu} ~
\Big\langle \exp \Big[ NK \mathbf{T}_{1} \sum_{\alpha}^{n} \Big\{  -\frac{1}{2}(m_{\alpha}^{1})^{2} + \sum_{i}^{N} m_{\alpha}^{1} (w_{i} \xi_{i}^{1} S_{i}^{\alpha})  \Big\} \Big] \nonumber\\
&&~ \times  \exp \Big[ NK \mathbf{T}_{1} \sum_{\mu = 2}^{p} \sum_{\alpha}^{n} \Big\{  -\frac{1}{2}(m_{\alpha}^{\mu})^{2} + \sum_{i}^{N} m_{\alpha}^{\mu} (w_{i} \xi_{i}^{\mu} S_{i}^{\alpha})  \Big\} \Big] \cdots \Big\rangle_{\xi} \nonumber\\
&=& \textrm{Tr} ~ \int \prod_{\mu}^{p} \prod_{\alpha}^{n} dm_{\alpha}^{\mu} ~
\exp \Big[ NK \mathbf{T}_{1} \sum_{\alpha}^{n} \Big\{  -\frac{1}{2}(m_{\alpha}^{1})^{2} + \sum_{i}^{N} m_{\alpha}^{1} (w_{i} \xi_{i}^{1} S_{i}^{\alpha})  \Big\} \Big] \times  \nonumber\\
&& \exp  \Big[ NK \mathbf{T}_{1} \sum_{\mu = 2}^{p} \sum_{\alpha}^{n} \Big\{  -\frac{1}{2}(m_{\alpha}^{\mu})^{2}\Big\} \nonumber \\
&+& \sum_{\mu = 2}^{p} \sum_{i}^{N} \ln \cosh \Big( N K \mathbf{T}_{1} \sum_{\alpha}^{n} m_{\alpha}^{\mu} (w_{i} S_{i}^{\alpha}) \Big) \Big] \cdots. ~~~
\end{eqnarray}

In the thermodynamic limit ($N \to \infty$), we expand $\ln \cosh (\cdots)$ to the second order of $m_{\alpha}^{\mu}$, so that
\begin{eqnarray}
\exp \bigg[ NK \mathbf{T}_{1} \sum_{\mu = 2}^{p} \sum_{\alpha}^{n} \Big\{  -\frac{1}{2}(m_{\alpha}^{\mu})^{2}\Big\} + \sum_{\mu = 2}^{p} \sum_{i}^{N} \ln \cosh \Big( NK \mathbf{T}_{1} \sum_{\alpha}^{n} m_{\alpha}^{\mu} (w_{i} S_{i}^{\alpha}) \Big) \bigg] \nonumber\\
\simeq \exp \bigg[ - \frac{1}{2}  NK \mathbf{T}_{1} \sum_{\mu = 2}^{p} \sum_{\alpha \beta}^{n} m_{\alpha}^{\mu} K_{\alpha \beta} m_{\beta}^{\mu} \bigg],
\label{eq:quad}
\end{eqnarray}
where $K_{\alpha \beta} \equiv \delta_{\alpha \beta} - K \mathbf{T}_{1} \sum_{i}^{N}  w_{i} S_{i}^{\alpha}  S_{i}^{\beta}$. Since Eq.~(\ref{eq:quad}) is quadratic in $m_{\alpha}^{\mu}$, the integral over $m_{\alpha}^{\mu}$ can be carried out using of the multi-variable Gaussian integral as follows:
\begin{eqnarray}
&&\int \prod_{\mu = 2}^{p} dm_{\alpha}^{\mu} ~\exp \Big[-\frac{1}{2} NK \mathbf{T}_{1} \sum_{\mu = 2}^{p} \sum_{\alpha \beta}^{n} m_{\alpha}^{\mu} K_{\alpha \beta} m_{\beta}^{\mu} \Big] \simeq  (\mathrm{det} \mathbf{K})^{-(p-1)/2} \nonumber\\
&=& \int \prod_{(\alpha \beta)}^{n} d q_{\alpha \beta}  ~\delta \Big( q_{\alpha \beta}  -  \sum_{i}^{N} w_{i} S_{i}^{\alpha}  S_{i}^{\beta} \Big)  \exp \Big( - \frac{p-1}{2}~\textrm{Tr}_{n} \ln \{ \mathbf{I} - K \mathbf{T}_{1} (\mathbf{I} + \mathbf{Q}) \}  \Big),
\label{eq:b14}
\end{eqnarray}
where $(\alpha \beta)$ denotes a summation over $\alpha$ and $\beta (\neq \alpha)$.
Here, we used the fact that the diagonal element $K_{\alpha \alpha}$ of the matrix $\mathbf{K}$ equals to $1 - K \mathbf{T}_{1}$ and the off-diagonal element becomes
\begin{eqnarray}
K_{\alpha \beta} = - K \mathbf{T}_{1} \sum_{i}^{N}  w_{i} S_{i}^{\alpha}  S_{i}^{\beta} \equiv - K \mathbf{T}_{1} q_{\alpha \beta} ,
\end{eqnarray}
which was expressed in terms of the spin glass order parameter $q_{\alpha \beta}$. The matrix $\mathbf{Q}$ consists of zeros along the diagonal elements and the off-diagonal elements $q_{\alpha \beta}$. $\textrm{Tr}_{n}$ is the trace of the $n \times n$ matrix. Using the Fourier representation of the $\delta$ function in Eq.~(\ref{eq:b14}) with integral variable $\tilde{r}_{\alpha \beta}$,
we can obtain the result:
\begin{eqnarray}
\langle \langle Z^{n} \rangle_{\xi} \rangle_{K} &=& \textrm{Tr} ~ \int  \prod_{\alpha}^{n} dm_{\alpha}^{1}
\prod_{(\alpha \beta)}^{n} d q_{\alpha \beta}  \prod_{(\alpha \beta)}^{n} d \tilde{r}_{\alpha \beta}
\exp \Big[  N K \mathbf{T}_{1} \sum_{\alpha}^{n} \Big\{  -\frac{1}{2}(m_{\alpha}^{1})^{2} + \nonumber \\
&&\sum_{i}^{N} m_{\alpha}^{1} (w_{i} \xi_{i}^{1} S_{i}^{\alpha})  \Big\} \Big] \nonumber\\
&&\times \exp \Big[ {\rm i} N K \mathbf{T}_{1} \sum_{(\alpha \beta)}^{n} \tilde{r}_{\alpha \beta}  \Big( q_{\alpha \beta}  - \sum_{i}^{N} w_{i} S_{i}^{\alpha}  S_{i}^{\beta} \Big) \Big] \nonumber \\
&& \times \exp \Big( - \frac{p-1}{2} ~\textrm{Tr}_{n} \ln \{(1 - K \mathbf{T}_{1})\mathbf{I} - K \mathbf{T}_{1} \mathbf{Q} \}  \Big).
\end{eqnarray}

The many-spin coupled problem is transferred to an average over a decoupled problem of single spins. For this simplification, we deal with inter-replica couplings in the single-spin problem. 
So we have
\begin{eqnarray}
\langle \langle Z^{n} \rangle_{\xi} \rangle_{K} = \textrm{Tr} ~ \int  \prod_{\alpha}^{n} dm_{\alpha}^{1}
\prod_{(\alpha \beta)}^{n} d q_{\alpha \beta}  \prod_{(\alpha \beta)}^{n} d \tilde{r}_{\alpha \beta}  ~\exp\{-N G(m_{\alpha}^{1}, q_{\alpha \beta} , \tilde{r}_{\alpha \beta} )\},
\end{eqnarray}
where
\begin{eqnarray}
G(m_{\alpha}^{1}, q_{\alpha \beta} , \tilde{r}_{\alpha \beta} ) &\equiv&
\frac{1}{2} K \mathbf{T}_{1} \sum_{\alpha}^{n} (m_{\alpha}^{1})^{2} - {\rm i} K \mathbf{T}_{1} \sum_{(\alpha \beta)}^{n} \tilde{r}_{\alpha \beta}  q_{\alpha \beta} \nonumber \\
&+& \frac{a}{2} ~\textrm{Tr}_{n} \ln \{(1 - K \mathbf{T}_{1})\mathbf{I} - K \mathbf{T}_{1} \mathbf{Q} \}- \ln \textrm{Tr}_{i} \exp (\tilde{\mathcal{H}}_{i})~~~
\end{eqnarray}
with
\begin{eqnarray}
\tilde{\mathcal{H}}_{i} \equiv N K \mathbf{T}_{1} \Big( \sum_{\alpha}^{n} m_{\alpha}^{1} (w_{i} \xi_{i}^{1} S_{i}^{\alpha})  -{\rm  i} \sum_{(\alpha \beta)}^{n} \tilde{r}_{\alpha \beta} (w_{i} S_{i}^{\alpha}  S_{i}^{\beta}) \Big).
\end{eqnarray}
Here we used $p-1 \simeq p = a N$ and the trace $\textrm{Tr}_{i}$ is now  at a single spin site.

In the thermodynamic limit ($N \to \infty$) the integrals can be
performed by the steepest descent method:
\begin{eqnarray}
\int dy ~\exp\{-NG(y)\} \simeq \int dy ~\exp \Big \{-NG(y_{0})-\frac{1}{2}NG''(y_{0})(y-y_{0})^{2}+\cdots \Big\},
\label{eq:b20}
\end{eqnarray}
where $G^{\prime}(y_{0})=0$ determines a saddle point $y_{0}$. The Gaussian
term can be ignored for $N \to \infty$, provided that $G''(y_{0}) > 0$.
Otherwise, the resulting integral diverges and the saddle point
procedure fails. Assuming $G''(y) > 0$, we replace $y$ with their stationary value. Therefore, we can define three order parameters as follows:
i) $m_{\alpha}^{\mu}$ represents the extent to which the $\mu$-th pattern of memory $\xi^{\mu}$ and the $\alpha$-th state of the system $S^{\alpha}$ overlap with each other;
\begin{eqnarray}
m_{\alpha}^{\mu} = \sum_{i}^{N} w_{i} \langle \xi_{i}^{\mu} S_{i}^{\alpha} \rangle .
\end{eqnarray}
ii) $q_{\alpha \beta} $ is the spin glass order parameter representing the extent to which the two states $\alpha$ and $\beta$ of the replica
overlap each other;
\begin{eqnarray}
q_{\alpha \beta}  = \sum_{i}^{N} w_{i} \langle S_{i}^{\alpha}  S_{i}^{\beta} \rangle .
\end{eqnarray}
iii) $r_{\alpha \beta} $ represents the extent to which the two different $m_{\alpha}^{\mu}$'s overlap each other;
 \begin{eqnarray}
r_{\alpha \beta} \equiv \frac{N}{p} \sum_{\mu = 2}^{p}  m_{\alpha}^{\mu} m_{\beta}^{\mu}
=\frac{1}{a} \sum_{\mu=2}^{p} q_{\alpha \beta}
= -\frac{2{\rm i}}{a K \mathbf{T}_{1}}\tilde{r}_{\alpha \beta}.~~~~~~
\end{eqnarray}
So, $r_{\alpha \beta} $ can be understood as the sum of the effects of non-retrieved patterns. Here, the average is evaluated through
$\langle A \rangle \equiv  \textrm{Tr}_{i} A \exp \tilde{\mathcal{H}}_{i} / \textrm{Tr}_{i} \exp \tilde{\mathcal{H}}_{i}$.

Therefore, the free energy becomes
\begin{eqnarray}
n\beta f &=& \frac{1}{2} K \mathbf{T}_{1} \sum_{\alpha}^{n} (m_{\alpha}^{1})^{2}  + \frac{1}{2} a K^{2} \mathbf{T}_{2} \sum_{(\alpha \beta)}^{n} r_{\alpha \beta} q_{\alpha \beta}  + \frac{a}{2} ~\textrm{Tr}_{n} \ln \{(1 - K \mathbf{T}_{1})\mathbf{I} - K \mathbf{T}_{1} \mathbf{Q} \} \nonumber \\ && -\ln \textrm{Tr}_{i} \exp (\tilde{\mathcal{H}}_{i}),
\label{eq:free_energy}
\end{eqnarray}
where
\begin{eqnarray}
\tilde{\mathcal{H}}_{i} = N K \mathbf{T}_{1} \sum_{\alpha}^{n} m_{\alpha}^{1} (w_{i} \xi_{i}^{1} S_{i}^{\alpha}) +  \frac{1}{2} a N K^{2} \mathbf{T}_{2} \sum_{(\alpha \beta)}^{n} r_{\alpha \beta} (w_{i} S_{i}^{\alpha} S_{i}^{\beta}).
\end{eqnarray}


\section{Replica-symmetric solutions}
\label{appen:rep_symm_sol}

Here we take the replica-symmetric (RS) assumption to set $q_{\alpha \beta} =q$ and $r_{\alpha \beta} =r$ for all $\alpha
\neq \beta$, and $m_{\alpha}^{1}=m$ for all $\alpha$. The RS solution ($m$, $q$, $r$) is the simplest one among several solutions obtained by the free energy (Eq.~(\ref{eq:free_energy})).
As we will be confirmed, the RS solution is stable for entire temperature regime but for very low temperature region near zero. Thus, using only the RS solution, we can analyze various characteristics of the Hopfield neural network.

Then the free energy is given by
\begin{eqnarray}
\beta f &=& \lim_{n \to 0} \frac{1}{n} \bigg[
\frac{1}{2} K \mathbf{T}_{1} n m^{2} + \frac{1}{2} K^{2} \mathbf{T}_{2} n(n-1) a r q + \frac{a}{2} ~\textrm{Tr}_{n} \ln \{(1 - K \mathbf{T}_{1})\mathbf{I} - K \mathbf{T}_{1} \mathbf{Q} \}\nonumber \\  && - \ln \textrm{Tr}_{i} \exp (\tilde{\mathcal{H}}^{\prime}) \bigg]
\label{eq:c1}
\end{eqnarray}
with  the effective Hamiltonian $\tilde{\mathcal{H}}^{\prime}$
\begin{eqnarray}
\tilde{\mathcal{H}}^{\prime} \equiv N K \mathbf{T}_{1} w_{i} m \xi^{1} \sum_{\alpha}^{n} S^{\alpha} + \frac{1}{2} N K^{2} \mathbf{T}_{2} w_{i} a r\sum_{(\alpha \beta)}^{n} S^{\alpha}S^{\beta}.~~~~~
\end{eqnarray}
To calculate the trace of the third term of Eq.~(\ref{eq:c1}), it should be noted that the eigenvectors of the matrix $\mathbf{Q}$ are, first, the uniform one $^{t}(1,1, \cdots, 1)$ and, second, the form of $^{t}(1, 0, \cdots, 0,  -1, 0, \cdots, 0)$.
So, the eigenvalue of the first eigenvector of the matrix $(1 - K \mathbf{T}_{1})\mathbf{I} - K \mathbf{T}_{1} \mathbf{Q}$ is $1- K \mathbf{T}_{1} - (n-1) K \mathbf{T}_{1} q$ (no degeneracy), and the eigenvalue of the second eigenvector is $1- K \mathbf{T}_{1} + K \mathbf{T}_{1} q$ (degeneracy $n-1$). Thus, in the limit $n \to 0$,
\begin{eqnarray}
\frac{1}{n}\textrm{Tr}_{n} \ln \{(1 - K \mathbf{T}_{1})\mathbf{I} - K \mathbf{T}_{1} \mathbf{Q} \}
&=& \frac{1}{n} \ln (1- K \mathbf{T}_{1} - (n-1) K \mathbf{T}_{1} q) \nonumber \\
&&+ \frac{n-1}{n} \ln (1- K \mathbf{T}_{1} + K \mathbf{T}_{1} q) \nonumber\\
&\stackrel{n \to 0}{\longrightarrow}& \ln (1- K \mathbf{T}_{1} + K \mathbf{T}_{1} q)  - \frac{K \mathbf{T}_{1} q}{1- K \mathbf{T}_{1} + K \mathbf{T}_{1} q}. ~~~~~
\end{eqnarray}

When we put $\xi^{1} = 1$, we finally obtain
\begin{eqnarray}
\beta f &=& \frac{1}{2} K^{2} \mathbf{T}_{2} a r(1-q) + \frac{1}{2} K \mathbf{T}_{1} m^{2}  + \frac{a}{2} \Big[ \ln (1- K \mathbf{T}_{1} + K \mathbf{T}_{1} q)  - \frac{K \mathbf{T}_{1} q}{1- K \mathbf{T}_{1} + K \mathbf{T}_{1} q} \Big] \nonumber \\
&& - \int \mathcal{D}z \frac{1}{N} \sum_{i=1}^{N}
\ln \Big[2 \cosh \eta_{i}(z) \Big].
\end{eqnarray}
where $\int \mathcal{D}z ~\cdots \equiv \frac{1}{\sqrt{2 \pi}} \int_{-\infty}^{\infty} dz ~\exp \{-\frac{1}{2}z^{2}\}~ \cdots $, and $\eta_{i}(z) \equiv K \mathbf{T}_{1} (z \sqrt{N w_{i} a r} + N w_{i} m)$.

We can determine  $m$, $r$, and $q$ by imposing the condition that $f$ resumes the stable extrema when they are the RS solutions. From this extremal condition, we can obtain the self-consistent equations of $m$, $q$, and $r$
as follows:
\begin{eqnarray}
m &=& \int \mathcal{D}z~ \sum_{i=1}^{N} w_{i} \tanh \eta_{i}(z) \label{eq:m}\\
q &=& \int \mathcal{D}z~ \sum_{i=1}^{N} w_{i} \tanh^{2} \eta_{i}(z) \label{eq:c6}\\
r &=& \frac{q}{ (1- K \mathbf{T}_{1} + K \mathbf{T}_{1} q)^{2}}. \label{eq:r}
\end{eqnarray}

The Almeida-Thouless (AT) line, i.e., the condition satisfying $G''(y_{0}) = 0$ in Eq.~(\ref{eq:b20}), is simply given by \cite{Amit87,Almeida78}
\begin{eqnarray}
(1- K \mathbf{T}_{1} + K \mathbf{T}_{1} q)^{2}  - K^{2} \mathbf{T}_{2} a \int \mathcal{D}z \sum_{i=1}^{N} N w_{i}^2~\textrm{sech}^{4} \eta_{i}(z) = 0. \label{eq:AT}
\end{eqnarray}
The AT lines for various $\gamma$ values locate in very low temperature region near zero so that even the SG and M phases become stable under replica symmetry.
Note that the dotted black line near zero temperature in each panel of Fig. 4 represents the AT line. Thus, the replica-symmetric solution is valid over almost the entire region.

We consider a particular case $T \to 0$ (i.e., $\mathbf{T}_{l} \to 1~~(l=1,2)$).
In this limit the ``$\tanh$" reduces to a step function
\begin{eqnarray}
\tanh \eta_{i}(z) \to \mathrm{sgn} (\eta_{i}(z)),
\end{eqnarray}
where $\eta_{i}(z) = K (z \sqrt{N w_{i} a r} + N w_{i} m)$.
Then the equation for $m$ becomes
\begin{eqnarray}
m &=& \int \mathcal{D}z~ \sum_{i=1}^{N} w_{i} ~ \mathrm{sgn} (\eta_{i}(z)) \nonumber\\
&=& \sum_{i=1}^{N} w_{i} ~ \frac{2}{\sqrt{2\pi}} ~ \int_{0}^{\sqrt{\frac{N w_{i}}{a r}} m} dz ~e^{-\frac{1}{2} z^{2}} \nonumber\\
&=& \sum_{i=1}^{N} w_{i} ~ \mathrm{erf} \Big( \sqrt{\frac{N w_{i}}{2 a r}} m \Big),
\end{eqnarray}
where $\mathrm{erf}(\cdots)$ means the error function.
The parameter $q$ approaches one, i.e., $q \to 1$ in the zero temperature limit.
A simple equation in this limit is readily obtained from $\partial (\beta f)/\partial r = 0$:
\begin{eqnarray}
NK (1-q) &=& \sum_{i=1}^{N} \sqrt{\frac{N w_{i}}{a r}} ~ \int \mathcal{D}z ~z ~\mathrm{sgn} (\eta_{i}(z)) \nonumber\\
&=& \sum_{i=1}^{N} \sqrt{\frac{N w_{i}}{a r}} ~ \frac{2}{\sqrt{2\pi}} ~ \int_{\sqrt{\frac{N w_{i}}{a r}} m}^{\infty} dz ~z ~e^{-\frac{1}{2} z^{2}} \nonumber\\
&=&\sum_{i=1}^{N} \sqrt{\frac{2 N w_{i}}{\pi a r}} ~\exp \Big( - \frac{N w_{i}}{2 a r} m^{2} \Big).
\end{eqnarray}
Therefore, at zero temperature, $m$, $q$ and $r$ are obtained as
\begin{eqnarray}
m &=& \sum_{i=1}^{N} w_{i} ~\mathrm{erf} \Big( \sqrt{\frac{N w_{i}}{2 a r}} m \Big)\label{eq:e4}\\
q &=& 1 - \frac{1}{N K} \sum_{i=1}^{N} \sqrt{\frac{2 N w_{i}}{\pi a r}} ~ \exp \Big(- \frac{N w_{i}}{2 a r} m^{2} \Big)\label{eq:e5}\\
r &=& \frac{q}{ (1- K + K q)^{2}}, \label{eq:e6}
\end{eqnarray}
which are Eqs.~(8-10) presented in the main text. Detailed explanations on these solutions are given in Fig 2 of the main text.

\section{Dependence of the phase boundaries on degree exponent}
\label{appen:phase}

As $\gamma$ is decreased, the $T-a$ phase diagram in Fig~1 of the main text undergoes drastic changes. Especially, the R phase introduces into the region of the SG phase, but it also raises the boundary of the P phase to a high-temperature region as $\gamma$ is decreased. As $\gamma$ approach 2.0, it is shown that the region of R phase becomes broader whereas the SG phase shrinks and eventually disappears. Such changes of the phase diagram in the $T-a$ space can be obtained by the leading term in the expansion of the right hand side of Eq.~(\ref{eq:c6}) with $m=0$, i.e., $q \simeq a r K^{2} \mathbf{T}_{2} X$ with $X \equiv N \sum_{i=1}^{N} w_{i}^{2}$, under the condition $N^{1/(\gamma -1)} \ll 1$. From this, we obtain the glass transition temperature to be $T_{g} \simeq 1/ K \tanh^{-1}(1/K (1+\sqrt{a X}))$. For the ER network ($\gamma \to \infty$) with $K=5$ and $a =0$, we obtain $T_{g} \simeq 1.013$ as shown in Fig~1(a) of the main paper. However, as $\gamma \to 2$, the condition $N^{1/(\gamma -1)} \ll 1$ cannot be fulfilled as $N \to \infty$, thus there exists no phase boundary between the P and SG phases and the R phase intrudes into the region between the two phases.

\section{Dependence of the error rate on degree exponent}
\label{appen:error_rate}

Fig~2 of the main text shows the error rate $n_{e} \equiv (1-m)/2$ as a function of storage rate $a$, obtained by Eqs.~(\ref{eq:e4}-\ref{eq:e6}) for various $\gamma$ values at zero temperature. As shown in Eq.(\ref{eq:e4}), $m$ has a $\gamma$ dependence in terms of $w_{i}$.
For the case $\gamma \to \infty$, $N w_{i}$ becomes unity and Eq.(\ref{eq:e4}) reduces to $m = \mathrm{erf} (m/\sqrt{2 a r})$, by which $m$ suddenly becomes zero when $a$ is larger than a specific value called the critical storage capacity $a_{c}=a_{c}(\gamma)$. For the case $\gamma \to 2 + \epsilon~~(\epsilon \ll 1)$, however, $N w_{i} \sim N^{1/(1+\epsilon)}i^{-1/(1+\epsilon)} \gg 1$ as $N \to \infty$, by which $m$ becomes nonzero even for sufficiently large value of $a$ and it is thus impossible to define the critical storage capacity $a_{c}$. Such a $\gamma$-dependence of $m$ determines the behavior of the error rate $n_{e}$, as shown in Fig~2 of the main text.

\section{Dependence of the phase diagrams on size $N$}

We remark that even though we obtained analytic formulae of Sec. III in the thermodynamic limit, the phase diagrams (Figs~1 and 2 of the main text) were obtained in finite systems. We find that when $\gamma$ is sufficiently large, $Np_i$ is not so sensitive to $N$ that the result of the ER case does not depend on $N$ seriously, as shown in Fig. 5(a). However, when $\gamma$ approaches 2.0, the R-phase region becomes wider as $N$ is increased, as shown in Fig. 5(b).

\section{Comparison between the Chung-Lu model and the static model}
\label{appen:static}

We show the phase diagram in the space $(T-a)$ and the error rate $n_e=(1-m)/2$ as a function of storage rate $a$ for the static model. First, Fig. 6(a) and (b) show the phase diagrams for the static model with (a) $\gamma=2.35$ and (b) $\gamma=2.01$, which correspond to the Figs~1(e) and (f) of the main paper for the CL model. These phase diagrams for the static model are similar qualitatively to those for the CL model, but the transitions temperature between P and R phases obtained for the static model is somewhat larger than the one for the CL model.

Fig. 7 is the plot of the error rate $n_{e} \equiv (1-m)/2$ vs storage rate $a$ at zero temperature for the two different $\gamma$ values for the static model. This figure corresponds to Fig~2(b) of the main paper for the CL model. Whereas the SG phase of the CL model remains on the axis $T=0$ when $\gamma \simeq 2.04$, as shown in Fig~2(b) of the main paper, the SG phase of the static model given in Fig. 7 remains on the axis $T=0$ when $\gamma \simeq 2.35$.

Fig. 8(a) and (b) show the comparison of the error rate for the two different models, the CL and the static models as a function of storage rate $a$ for different degree exponent values (a) $\gamma=2.01$ and (b) $2.04$ at $T=0$. These figures enable us to see qualitatively how the error rate depends on the degree-degree correlation. We recall that the disassortative degree-degree correlation is present for the static model in the region $2 < \gamma <3$, but absent for the CL model. One can see that the degree-degree correlation reduces somewhat the error rate. Accordingly one may infer that the error rate of realistic brain networks, in which the degree-degree correlation is present, should be smaller than those obtained from the CL model analytically.

\section{Hopfield model simulations on real neural networks}
\label{appen:real_nets}

We check the error rate $n_{e}$ of the Hopfield model on several real neural networks. For this purpose, we chose three real neural networks: the networks for Macaque monkey 1 (fve30; $N = 30$, $L=311$) \cite{Felleman91}, Macaque monkey 2 (macaque47; $N=47$, $L=505$) \cite{Honey07} and cat (CIJctx; $N=52$, $L=818$) \cite{Scannell99}.  Those networks are constructed based on the connectivity network data sets~\cite{bctnet}. For the cat network, all links with nonzero elements (1, 2 and 3) of the adjacency matrix were regarded as connected.

Our simulations are performed following the random updating method of the Hopfield model on those real neural networks, which is the same as the method already used in Sec. 1(\nameref{appen:simulation}).
Here, the error rate $n_{e}(\equiv (1-m)/2)$ is calculated using the formula $m = (1/N) \sum_{i}^{N} \xi_{i}^{1} S_{i}(t\to \infty)$ with $\mu=1$. We perform $10^4$ different runs with $10^4$ different ensemble of $\{\xi_i^{\mu}\}$, and then obtain the error rate $n_e$ over those ensemble.

 Fig. 9 shows the degree distributions (left column) and the error rates (right column) of those three different real neural networks. The degree distributions of three networks are heavy-tailed, but the degree exponents cannot be determined precisely due to the small system sizes of those networks. To compare the error rates obtained from the simulations with the ones from the theory, we draw the  error rates obtained from both methods together in the right column of Fig. 9. Indeed, with appropriate weight $w_i(\gamma)$ near $\gamma = 2.0$, the two error rates behave similarly to each other as a function of the storage rate $a$.




\nolinenumbers

%
%
%


\newpage

\nolinenumbers

%
%
%


\begin{figure*}
\centering
\includegraphics[width=0.88\textwidth]{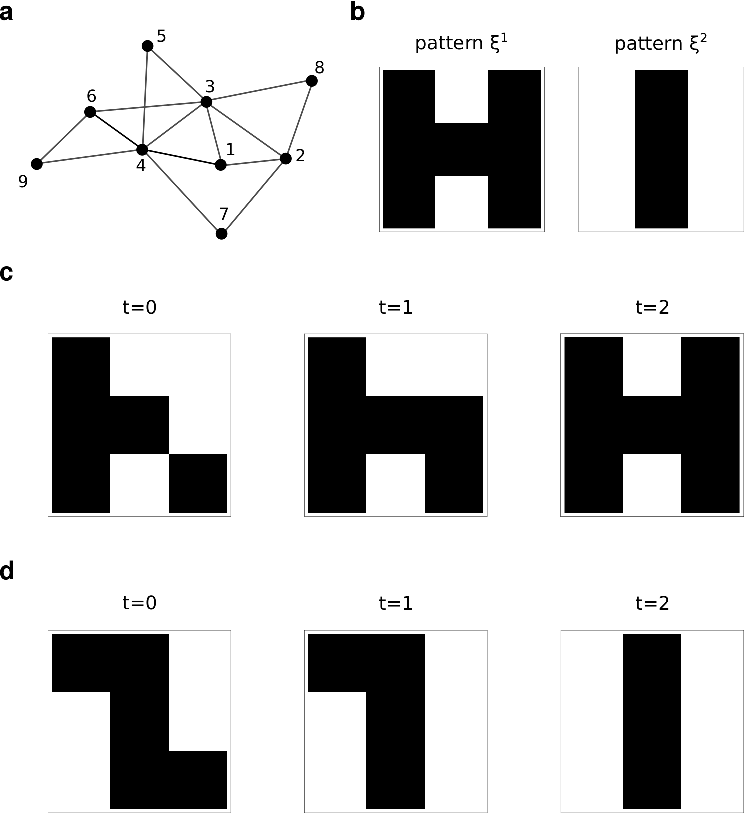}
\caption{A simple example of simulation of the Hopfield model. (a) A network of size $N=9$ and mean degree $K=\frac{5}{3}$. (b) Two patterns $\xi^1$ and $\xi^2$ are encoded by the Hebbian rule. In (c) and (d), we start from two random patterns. After some updates using Eq.(14), the patterns $\xi^1$ and $\xi^2$ are retrieved.}
\end{figure*}

\begin{figure*}
\centering
\includegraphics[width=0.88\textwidth]{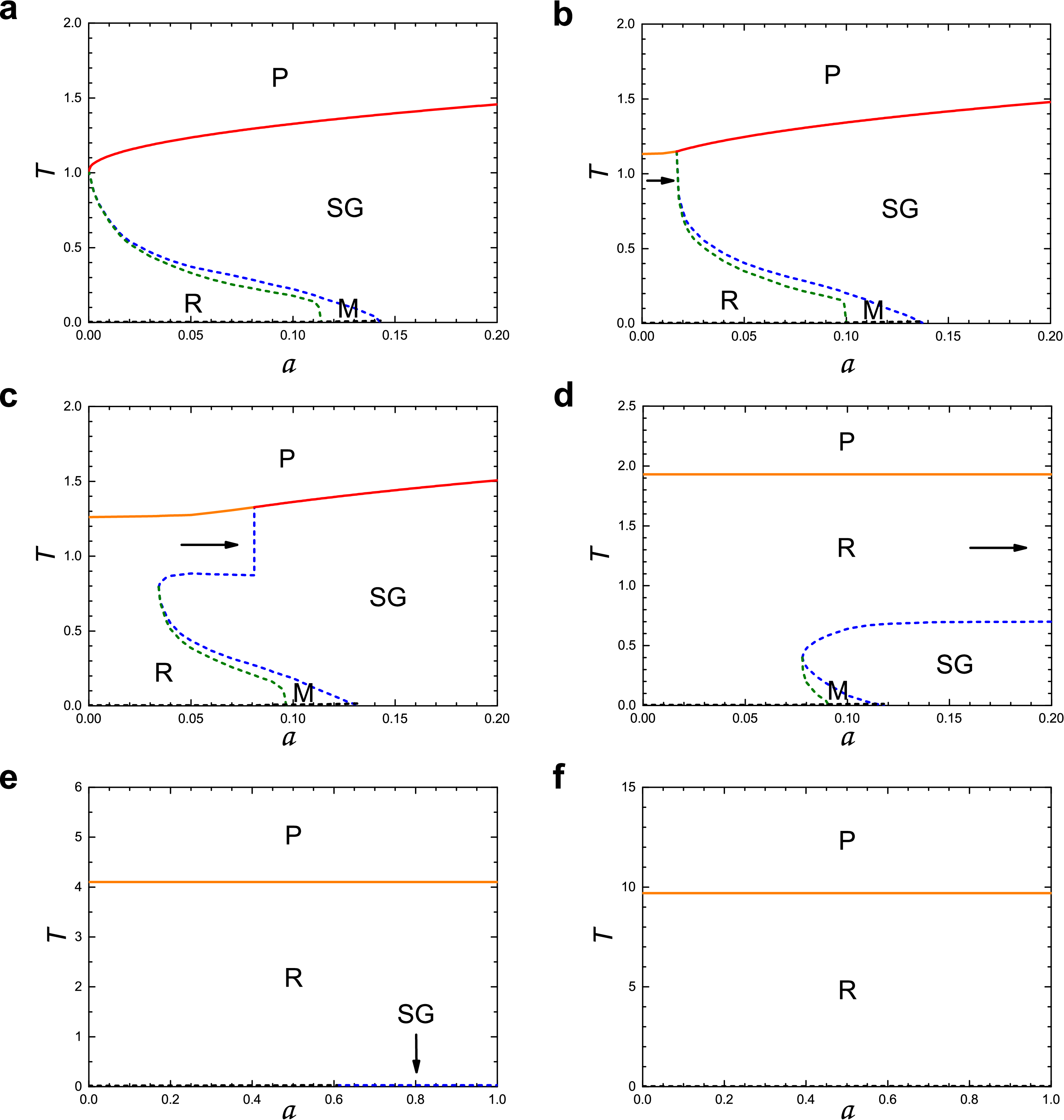}
\caption{Phase diagrams shown in Fig 1 of the main paper are redrawn with the Almeida-Thouless(AT) line (Eq.(45)).
Note that the dotted black line near zero temperature in each panel represents the AT line. Thus, we check that the replica-symmetric solution is valid over almost the entire region in the plane of $(T, a)$.}
\end{figure*}

\begin{figure*}
\centering
\includegraphics[width=0.88\textwidth]{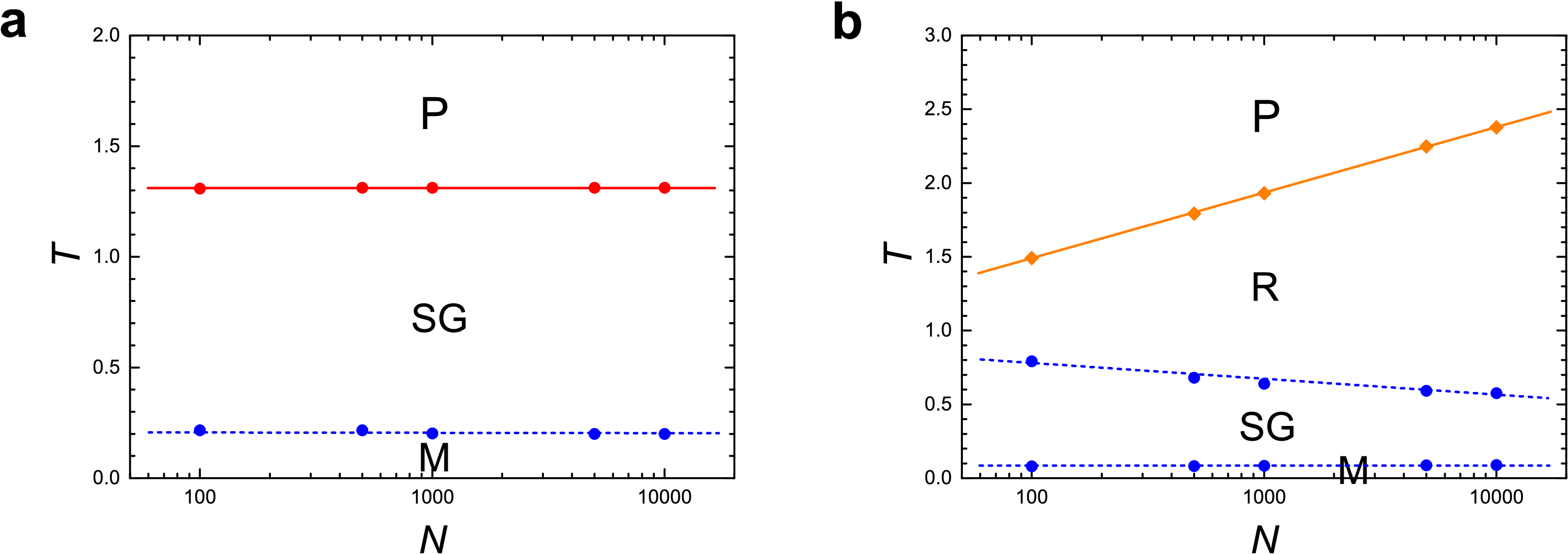}
\caption{Phase diagrams of the Hopfield model in the plane of $(T, N)$ on the Chung-Lu model of SF networks with (a) $\gamma=5.00$ and
(b) $\gamma=3.00$. Here, $a$ is fixed as 0.1.}
\end{figure*}

\begin{figure*}
\centering
\includegraphics[width=0.88\textwidth]{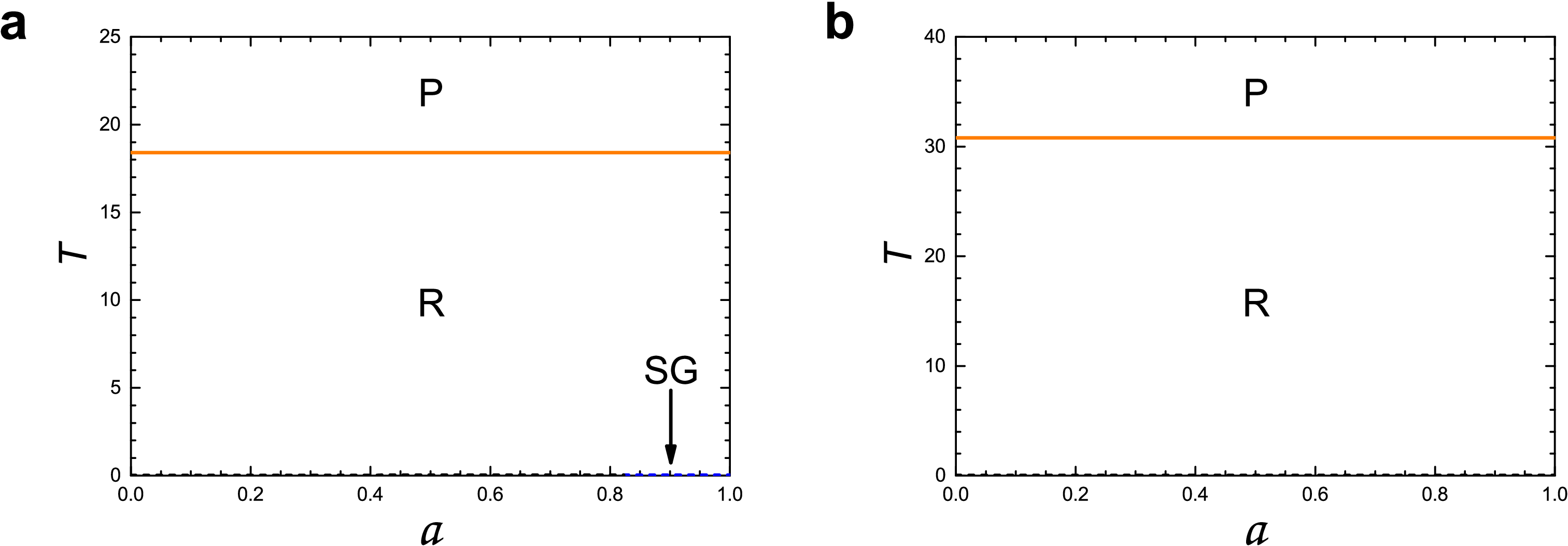}
\caption{Phase diagrams of the Hopfield model in the plane of $(T, a)$ on the static model of SF networks with (a) $\gamma=2.35$ and
(b) $\gamma=2.01$. Those phase diagrams correspond to Fig~1(e) and (f) of the main paper for the CL model, respectively. {\bf P} represents  paramagnetic phase, {\bf SG} spin glass phase, and {\bf R} retrieval phase.  The SG phase remains only on the axis $T=0$ when $\gamma = \gamma_{c} \simeq 2.35$, then the R phase spans the entire region of $a$ at the lower temperatures. Thus, the static model shows better memory retrieval than the CL model. To obtain the phase boundary, numerical calculations were performed for the static model with plugging $N = 1000$ and $K = 5.0$
into the formulas. Solid and dotted curves indicate the second-order and the first-order transitions, respectively. Note that black dotted line in each panel near the $T=0$ line represents the AT line (Eq.(45)). Thus, replica-symmetric solution is valid over almost the entire region of the phase space.}
\end{figure*}

\begin{figure*}
\centering
\includegraphics[width=0.88\textwidth]{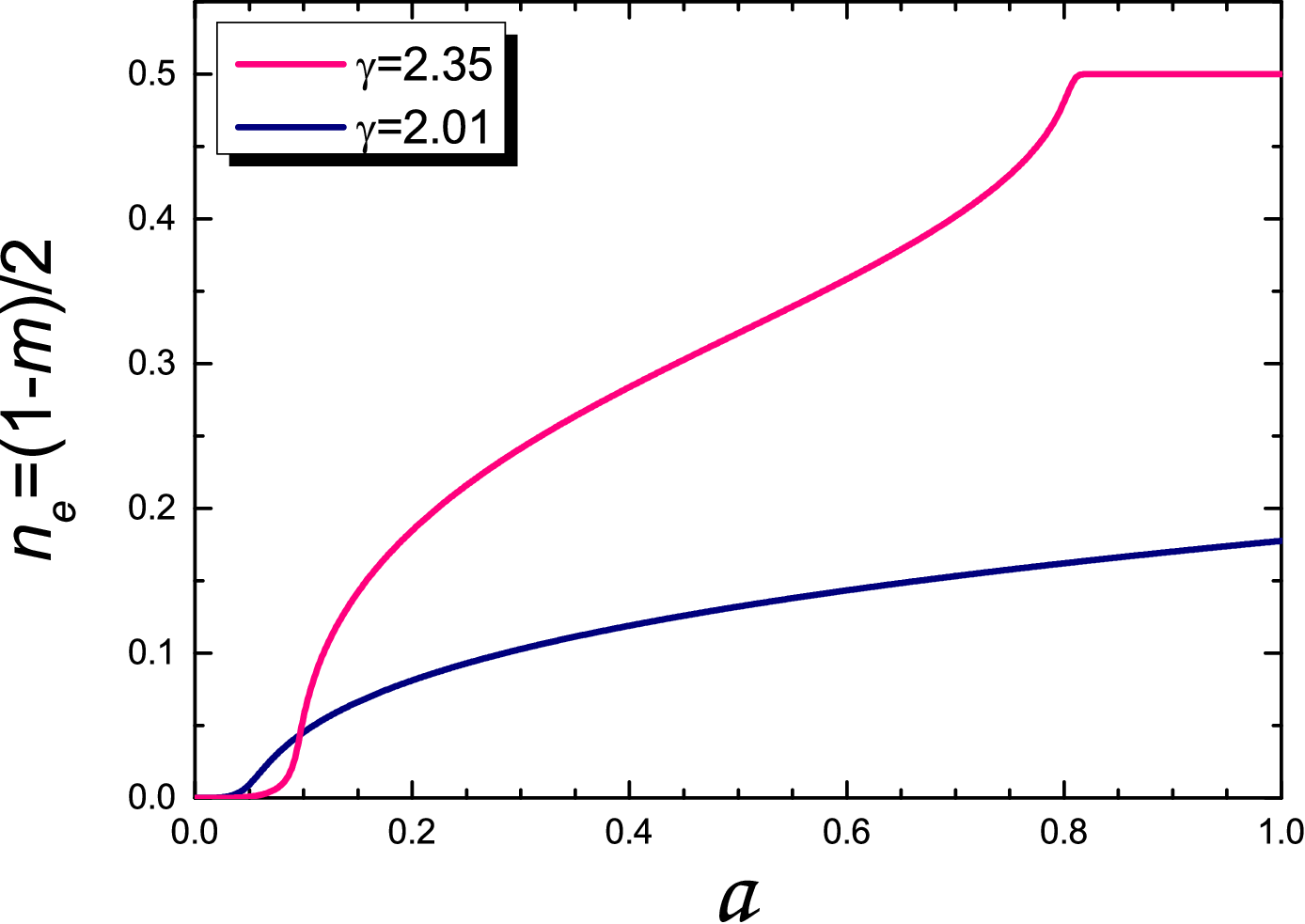}
\caption{Plot of the error rate $n_{e} \equiv (1-m)/2$ vs storage rate $a$ for $\gamma=2.01$ and $2.35$ for the static model at $T=0$, which corresponds to Fig~2(b) of the main paper for the CL model. Here, numerical values are obtained  using $N = 1000$ and $K = 5.0$.}
\end{figure*}

\begin{figure*}
\centering
\includegraphics[width=0.88\textwidth]{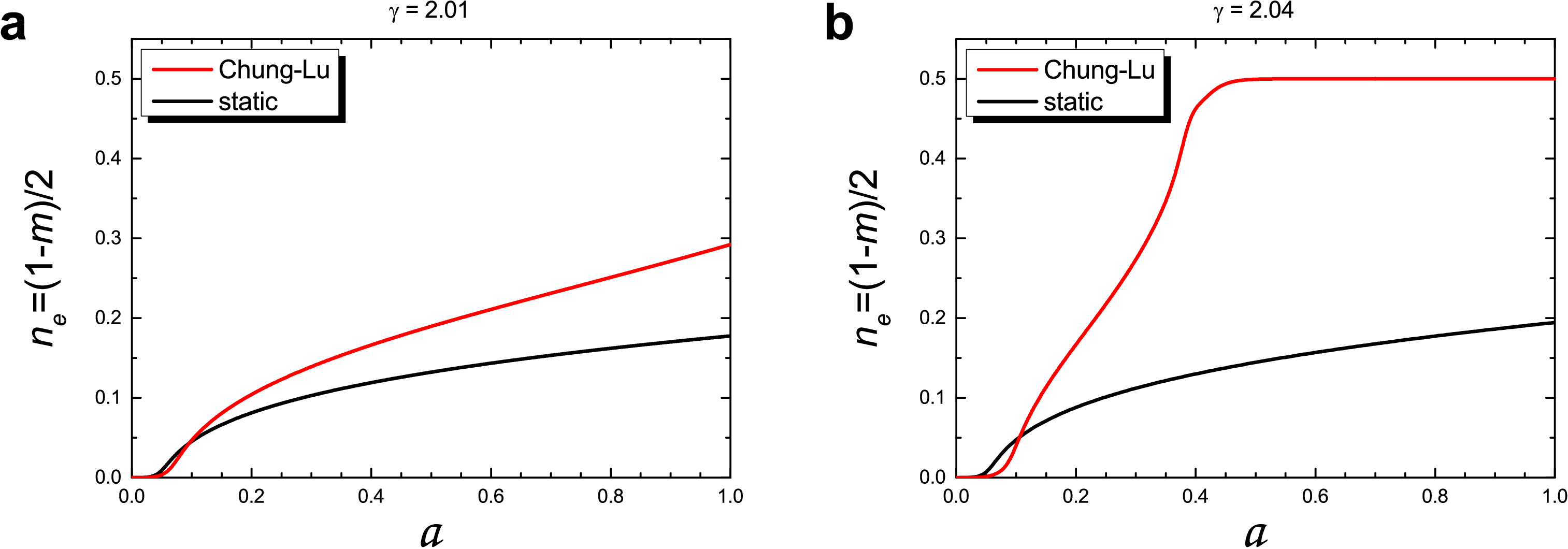}
\caption{Plot of the error rate $n_{e}$ vs storage rate $a$ for (a) $\gamma=2.01$ and (b) $\gamma=2.04$ for the CL and the static model at $T=0$. These figures provide the comparison of the error rate between for the CL model and for the static model. Here, numerical values are obtained using $N = 1000$ and $K = 5.0$.}
\end{figure*}

\begin{figure*}
\centering
\includegraphics[width=0.88\textwidth]{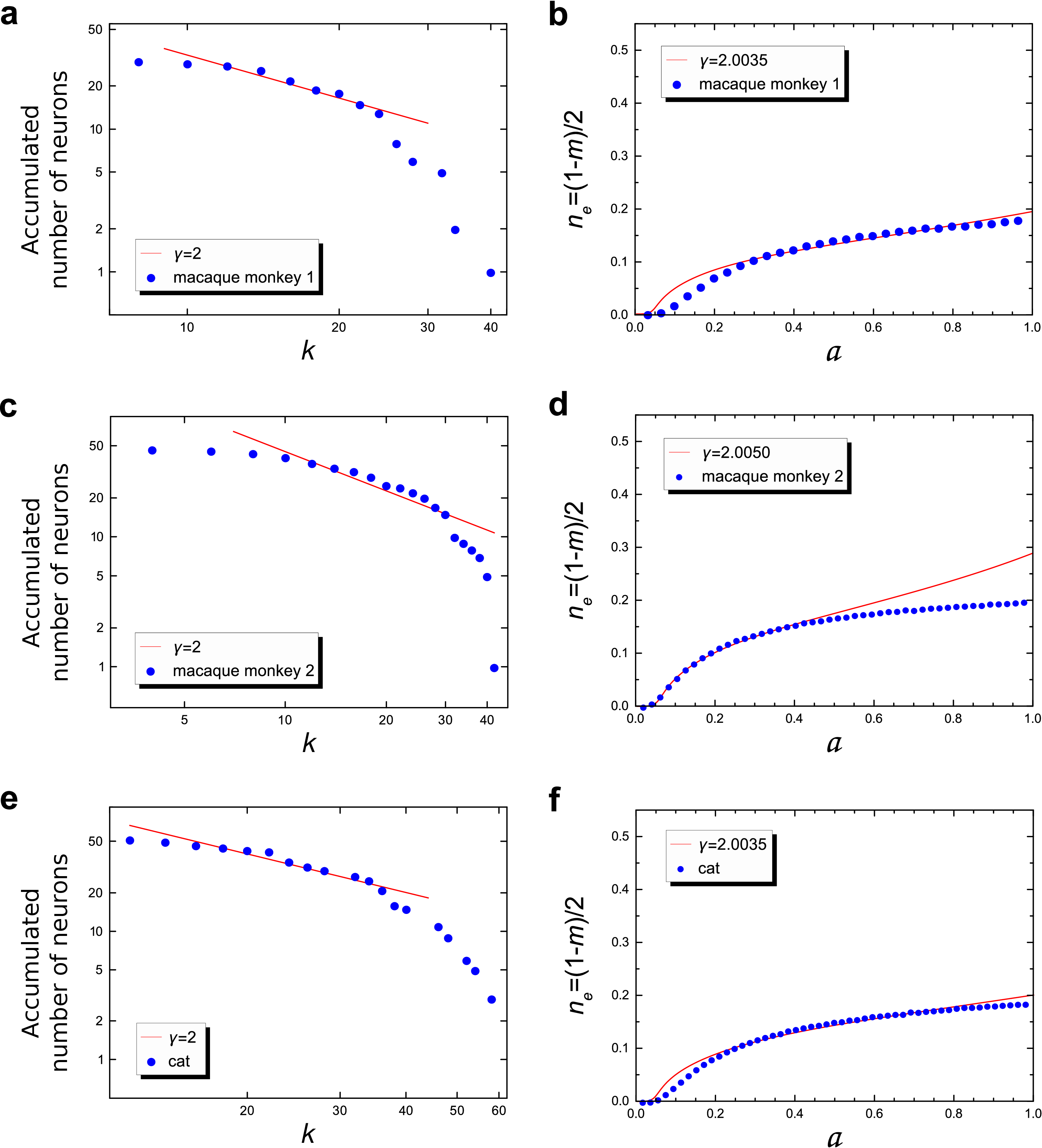}
\caption{Plots of the number of neurons (nodes) with degree larger than $k$ vs degree $k$ of the real neural networks (left column) and of their  error rates $n_{e}$ vs storage rate $a$ (right column): (a) and (b) for the visual cortex network of macaque monkey, with $N=30$ and $L=311$. (c) and (d) for the corticocortical connectivity network in the visual and sensorimotor area of macaque monkey, with $N=47$ and $L=505$. (e) and (f) for the cortex network of cat, with $N=52$ and $L=818$. The red lines of (a), (c) and (e) are guidelines with slope $-1$ for eye drawn to compare the data points with the scale-freeness of $\gamma = 2.0$. The red curves of (b), (d) and (f) are drawn to compare the simulation data ($\bullet$) with the analytic solution (red curve) using Eq.(49) under the same conditions of $N$ and $L$. Here, the $\gamma$ values we used for $w_{i}$ in Eq.(49) were 2.0035 (b), 2.0050 (d), and 2.0035 (f), respectively.}
\end{figure*}

\end{document}